\documentclass[12pt]{article}
\usepackage[left=2.5cm,top=2.5cm,right=2.5cm,bottom=1.5cm]{geometry}
\usepackage{graphicx}
\usepackage{latexsym}
\usepackage{epstopdf}
\usepackage{amsmath}
\usepackage{xcolor}
\DeclareGraphicsExtensions{.eps}
\usepackage{cite}
\begin{document}
\begin{center}
\large{\bf{A Novel Cosmic Framework of Interdependent Dark Matter and Holographic Dark Energy within the Bianchi Type-V Universe}}\\
\vspace{10mm}
\normalsize{ Gunjan Varshney$^1$, Anirudh Pradhan$^2$, Nasr Ahmed$^3$, and Vansh Mittal$^4$  }\\
\vspace{5mm}

\normalsize{$^{1}$ Department of Applied Sciences, HMR Institute of Technology and Management, Hamidpur, New Delhi 110036, India}\\
	\vspace{2mm}
\normalsize{$^{2}$ Centre for Cosmology, Astrophysics and Space Science (CCASS), GLA University,\\
	Mathura-281 406, Uttar Pradesh, India}\\
	\vspace{2mm}
\normalsize{$^{3}$ Astronomy Department, National Research Institute of Astronomy and Geophysics, Helwan, Cairo, 
	Egypt}\\
	\vspace{2mm}
\normalsize{$^{4}$ 11th Grade, Carmel High School, Indiana, USA}\\
\vspace{2mm}
$^1$ E-mail:gunjanvarshney03@gmail.com \\
\vspace{2mm}
$^2$ E-mail:pradhan.anirudh@gmail.com\\
	\vspace{2mm}
$^3$ E-mail:nasr.ahmed@nriag.sci.eg \\
	\vspace{2mm}
$^4$ E-mail:vanshmittal0427@gmail.com \\

\end{center}
\vspace{3mm}
\begin{abstract}
We study the anisotropic and homogeneous Bianchi type-V Universe with holographic dark energy (HDE) and interacting dark matter (DM). The solution for the field equations have been obtained for a certain form of the deceleration parameter. As for $\Lambda$CDM, we show that the coincidence problem disappears for a specific choice of the dark matter-holographic dark energy interaction. In this study, the observational data combination of OHD and JLA (Yu et al., Astrophys. J. 856 (2018) 3), where $q_0$ = $-0.52$ and $H_0$ = $69.2$, have been taken into consideration. We also found that the anisotropy of the expansion achieves isotropy after some finite time. We have also explained the physical and geometrical aspects of the model. The physical acceptability and stability of the model have been examined.
\end{abstract}
Keywords: Dark matter, Holographic dark energy, Bianchi type-V, Anisotropy, Homogeneous.
\section{Introduction}

In our knowledge of the Universe, we have witnessed incredible progress in the past few decades. It is assumed that two mystifying components named dark energy (DE) and dark matter (DM) compose 95\% of the Universe. The mysterious nature of both components depicts some introductory queries and implies a completely unknown physics, which is yet to be explored. Individually, 25\% of the Universe's whole energy density is comprised of DM. The presence of DM is firmly based upon the astrophysical observance in a comprehensive range of scales, whereas the nature of DM is yet obscure. In exact, our Universe's more mystifying component is DE. Various cosmological observances verified the existence of this kind of energy \cite{refaa1, refaa2, refaa3, refaa4}. However, the identical behavior of DE is always an open issue that needs additional investigations. The foremost and easy candidate of DE is the cosmological constant, despite confronting two severe theoretical concerns, likely fine-tuning and coincidence. Therefore, by examining the dynamics of distinct DE models and modifying Einsteins-Hilbert's action of general relativity that directs to the hypothesis of modified gravity, one can handle the DE issues. The DE models can be explored through the equation of state (EoS) parameter $\omega = \frac{p}{\rho}$ of the cosmological constant, where $p$ represents pressure and $\rho$ represents energy density.\\

Nevertheless, the essence of such dark energy is still slightly doubtful. 
The researchers looked out the description in a vast range of physical phenomena together with exotic fields, unique geometric forms of spacetime, cosmological constant, and new structures of the geometric equation, etc \cite{refaa5, refaa6, refaa7, refaa8, refaa9, refaa10}. Lately, the holographic principle (HP) \cite{holo1} has promoted a new model that is placed ahead by the researchers to define dark energy (DE) \cite{refaa11, refaa12}. 
 According to HP's analysis, a physical system's degrees of freedom increase as its bounding area increases. In light of this, the researchers of \cite{refaa13} assumed that an infrared (IR) cut-off and an ultraviolet (UV) cut-off are related because the black hole's formation imposes a constraint that sets an upper bound on the zero-point energy density.
Taking this into account, Hsu and Li \cite{refaa11, refaa12, refaa14} suggested that one can consider this energy density as holographic dark energy (HDE) fulfilling $\rho_{D} = 3 c^{2} / L^{2}$, where L is an IR cut-off in units $M^{2}_{p} = 1$, and $c^{2}$ is a constant. In \cite{refaa11, refaa12}, Li studied three selections for $L$: {the length scale} that are assumed to deliver the IR cut-off as the event horizon, the particle horizon, and the Hubble radius. Li showed that one can acquire the expected accelerating Universe and receive the accurate equation of state (EoS) parameter of DE only by remembering $L$ with the radius of the future event horizon. 
Aside from various uses of HP, the HDE model provides a novel example for the importance of holography in cosmology \cite{refaa15, refaa16, refaa17, refaa18, refaa19, refaa20,refaa21, refaa22, refaa23, refaa24, refaa25, refaa26}. The observational data and the HDE model are compatible \cite{refaa27, refaa28, refaa29}.
Wang et al. \cite{refA7} have discussed holographic dark energy (HDE), a theoretical approach that applies the holographic principle (HP) to the dark energy (DE) problem. The authors \cite{refA7} explored the relationship between DM and HDE, including topics like spatial curvature, neutrino, instability of perturbation, time-varying gravitational constant, inflation, black hole, and big rip singularity, from both theoretical and observational perspectives. Li et al. \cite{refA8} studied the cosmological implications of holographic dark energy (HDE) and interacting holographic dark energy (IHDE) models using CMB, DESI BAO, and SN data. Trivedi et al. \cite{refA9} have recently proposed a new holographic dark energy scenario called ``Fractional Holographic Dark Energy" (FHDE).  This model expands the standard framework of HDEs by including special elements from fractional calculus and the fractional Wheeler-De Witt equation, which have lately been employed, for example, in cosmological contexts. A large number of dark energy models, including entropic DE models, were discovered to be equivalent to the generalized HDE, with the associated cut-offs defined in terms of the particle horizon and its derivatives or the future horizon and its derivatives \cite{refA10}.
Lucianov \cite{refA11} investigated the quantum entropy-driven modifications to holographic dark energy in f(G, T) gravity. This technique broadens the usual HDE model by replacing the basic Bekenstein-Hawking entropy with Barrow entropy, which captures quantum gravitational adjustments to the geometry of black hole horizons. Huang et al. \cite{refA12} have recently investigated the holographic inflation and holographic dark energy from entropy of the anti-de Sitter black hole. This paper \cite{refA12} uses the Hubble horizon as the IR cutoff to study holographic and slow-roll inflation within the model. An interacting Holographic dark energy (HDE) with different infra-red (IR) cutoffs (Hubble horizon and future event horizon) is studied in the background dynamics of a flat FLRW universe with gravitational particle creation effects via various particle creation rates \cite{refA13}. Holographic dark energy models and their behaviors within the framework of f(Q,C) gravity theory is discussed in \cite{refA14}. \\

Latterly, Bianchi types are performing an essential part in observational cosmology therefore, the data of WMAP \cite{refaa30, refaa31, refaa32} appears to require an expansion with a positive cosmological constant in the classical cosmological model, which holds a similarity to the Bianchi analysis \cite{refaa33, refaa34, refaa35, refaa36, refaa37, refaa38}. As stated, the cosmic anisotropic geometry should be reached despite the inflation, which conflicts with the models of generic inflation \cite{refaa39, refaa40, refaa41, refaa42, refaa43, refaa44, refaa45} signifying a non-trivial isotropization past of the Universe. It is usually considered that this has experienced a term of exponential evolution to describe the flatness and uniformity of the currently surveyed Universe \cite{refaa39, refaa41, refaa42, refaa43}. Primarily the frame of the isotropic and homogeneous FLRW cosmology expresses the evolution of the Universe. All the reasons are purely technical where the field equations are simple and analytical solutions can be presented. To believe in the former inflationary era, there are no compelling physical reasons yet. 
Various researchers \cite{refaa46,refaa47,refaa48,refaa49} have explored certain issues of anisotropic models and discovered that the FLRW model stands intact for the expected method even when large anisotropies were present before the inflationary era. Therefore, the Bianchi model of anisotropy has evolved as an academic interest. Out of this, We have considered Bianchi type-V model in the current research. Goswami et al. \cite{refA15} constrained the cosmological parameters of the Bianchi type V universe by comparing the model under discussion to recent observational H(z) and Pantheon data and obtained the present value of deceleration parameter and age of universe as $-0.588022$ and $13.329$ Gyrs respectively. Yadav et al. \cite{refA16}  studied an anisotropic model of a transitioning universe with a hybrid scalar field under Brans-Dicke gravity, finding that the universe in the resultant model verifies Mach's principle and is consistent with current data. Goswami et al. \cite{refA17} studied a bulk viscosity-dominated accelerating expansion of a spatially homogeneous but anisotropic Universe in terms of kinematic parameters such as the Hubble parameter, deceleration parameter, and jerk parameter. They \cite{refA17} also conducted an analysis using Pantheon compilation data from 1048 SNIa apparent magnitude measurements, including 276 SNIa in the low-redshift range $0.3 \leq z \leq 0.65$. \\

An impressive law of variation for the average Hubble's parameter in Bianchi type-V cosmology has been suggested in \cite{refaa50,refaa51,refaa52}, resulting in a constant deceleration parameter value. These Hubble's parameter variation rules not only agree with observations, but also approach near accuracy for slowly time-varying DP models. The laws give a comprehensive framework of scale parameters to manage the Bianchi type-V Universe and also provide an explanation of the accelerating and decelerating mechanisms of the Universe's evolution. The works provide a general analysis of the constant DP models in different frameworks \cite{refaa50,refaa51,refaa52}. Under the assumption that there is perfect fluid or ordinary matter in the universe, researchers primarily looked at models with constant DP. However, ordinary stuff is insufficient to explain the dynamics of the accelerating Universe.\\

Motivated by the aforementioned conversations, this study aims to investigate the cosmological model in the context of interacting DM and HDE in the Bianchi type-V Universe. The following are the research's broad strokes: The field and metric equations are presented in Sec. 2, and the cosmological solutions are covered in Sec. 3. The physical acceptability of the solutions is covered in Sec. 4, which is broken down into two subsections. Subsection 4.1 explains the sound of speed and energy conditions are illustrated in subsection 4.2. The last Sec. 5 summarizes the discussion of the research.

\section{Basic field equations}

One way to express the Bianchi type-V line element is
\begin{eqnarray}
\label{1}
ds^{2}=dt^{2}-\alpha_{1}^{2}(t) dx^{2}-\alpha_{2}^{2}(t) e^{-2\gamma x} dy^{2}-\alpha_{3}^{2}(t) e^{-2\gamma x} dz^{2},
\end{eqnarray}
where $\alpha_{1}(t), \alpha_{2}(t)$ and $\alpha_{3}(t)$ are the cosmic scalar factors and $\gamma \neq 0$ is an arbitrary constant. In natural limits $(8\pi G = 1~and~ c = 1)$, Einstein equations are
\begin{eqnarray}
\label{2}
R_{ij}-\frac{1}{2} g_{ij} R = -\Big(^{dm}T_{ij}+^{de}T_{ij}\Big),
\end{eqnarray}
where
\begin{eqnarray}
\label{3}
^{dm}T_{ij}=\rho_{dm}u_{i}u_{j}, \nonumber \\
 ^{de}T_{ij} = \big(\rho+~~p\big)u_{i}u_{j}-g_{ij}~~p,
\end{eqnarray}
are energy-momentum tensors for DM (~pressure-less i.e $\omega_{dm} = 0$~) and HDE respectively. 
In this case, the energy density (ED) of DM is $\rho_{dm}$. The ED and pressure of HDE are denoted by $\rho$ and $p$.
With the aid of equation (3), the Einstein's field equation (2) for the metric (1) in co-moving coordinate systems can be expressed as
\begin{eqnarray}
\label{4}
\frac{\dot{\alpha_{1}} \dot{\alpha_{2}}}{\alpha_{1} \alpha_{2}} + \frac{\dot{\alpha_{2}} \dot{\alpha_{3}}}{\alpha_{2} \alpha_{3}} + \frac{\dot{\alpha_{3}} \dot{\alpha_{1}}}{\alpha_{3} \alpha_{1}} - \frac{3\gamma^{2}}{\alpha_{1}^{2}} = \rho + \rho_{dm},
\end{eqnarray}
\begin{eqnarray}
\label{5}
\frac{\ddot{\alpha_{2}}}{\alpha_{2}} + \frac{\ddot{\alpha_{3}}}{\alpha_{3}} + \frac{\dot{\alpha_{2}}\dot{\alpha_{3}}}{\alpha_{2} \alpha_{3}}  - \frac{\gamma^{2}}{\alpha_{1}^{2}} = -~~p,
\end{eqnarray}
\begin{eqnarray}
\label{6}
\frac{\ddot{\alpha_{1}}}{\alpha_{1}} + \frac{\ddot{\alpha_{3}}}{\alpha_{3}} + \frac{\dot{\alpha_{1}}\dot{\alpha_{3}}}{\alpha_{1} \alpha_{3}}  - \frac{\gamma^{2}}{\alpha_{1}^{2}} = -~~p,
\end{eqnarray}
\begin{eqnarray}
\label{7}
\frac{\ddot{\alpha_{1}}}{\alpha_{1}} + \frac{\ddot{\alpha_{2}}}{\alpha_{2}} + \frac{\dot{\alpha_{1}}\dot{\alpha_{2}}}{\alpha_{1} \alpha_{2}}  - \frac{\gamma^{2}}{\alpha_{1}^{2}} = -~~p,
\end{eqnarray}
\begin{eqnarray}
\label{8}
 \frac{\dot{\alpha_{2}}}{\alpha_{2}} +  \frac{\dot{\alpha_{3}}}{\alpha_{3}} = 2  \frac{\dot{\alpha_{1}}}{\alpha_{1}},
\end{eqnarray}
where the derivative with respect to time $t$ is represented by an overhead dot ($^{.}$).
When we integrate equation (8), we obtain
\begin{eqnarray}
\label{9}
\alpha_{1}^{2} = \lambda\alpha_{2}\alpha_{3},
\end{eqnarray}
where the integration constant is $\lambda$. In order to maintain generality, we can assume $\lambda = 1$. The average scale factor (SF) $a$ and volume $V$ are provided by
\begin{eqnarray}
\label{10}
V = a^{3} = \alpha_{1}\alpha_{2}\alpha_{3}.
\end{eqnarray} 
The definition of the mean Hubble parameter $H$ is
\begin{eqnarray}
\label{11}
H = \frac{\dot{a}}{a}=\frac{1}{3}\big(H_{x}+H_{y}+H_{z}\big),
\end{eqnarray}
where $H_{x} = \frac{\dot{\alpha_{1}}}{\alpha_{1}}$,~$H_{y} = \frac{\dot{\alpha_{2}}}{\alpha_{2}}$~and ~ $H_{z} = \frac{\dot{\alpha_{3}}}{\alpha_{3}}$ are the directional Hubble parameters, corresponding to the directions of x, y, and z axes respectively. 
The definition of the deceleration parameter (DP) $q(t)$ is
\begin{eqnarray}
\label{12}
q=-\frac{a\ddot{a}}{\dot{a}^{2}}.
\end{eqnarray}
The mean anisotropy expansion parameter $(\Delta)$, the expansion scalar $\theta$  and the shear scalar $\sigma^{2}$ are defined as
\begin{eqnarray}
\label{13}
\Delta = \frac{1}{3} \sum_{i=1}^{3}\left(\frac{H_{i}-H}{H}\right)^{2},
\end{eqnarray}
\begin{equation}
\label{14}
\theta = \left(\frac{\dot \alpha_{1}}{\alpha_{1}}+\frac{\dot \alpha_{2}}{\alpha_{2}}+\frac{\dot \alpha_{3}}{\alpha_{3}}\right) = 3 H,
\end{equation}
\begin{equation}
\label{15}
\sigma^2 = \frac{1}{2} \left(\sum_{i=1}^{3} H_i ^2 - \frac{\theta^2}{3}\right).
\end{equation}
Subtracting Eq.(5) from (6), Eq(6) from (7), Eq (5) from (7), and using Eq. (10), we get
\begin{eqnarray}
\label{16}
  \left(\frac{\dot{\alpha_{1}}}{\alpha_{1}} - \frac{\dot{\alpha_{2}}}{\alpha_{2}}\right) \frac{\dot{V}}{V} + \frac{d}{dt}\left(\frac{\dot{\alpha_{1}}}{\alpha_{1}} - \frac{\dot{\alpha_{2}}}{\alpha_{2}}\right) = 0,
\end{eqnarray}
\begin{eqnarray}
\label{17}
  \left(\frac{\dot{\alpha_{2}}}{\alpha_{2}} - \frac{\dot{\alpha_{3}}}{\alpha_{3}}\right) \frac{\dot{V}}{V} + \frac{d}{dt}\left(\frac{\dot{\alpha_{2}}}{\alpha_{2}} - \frac{\dot{\alpha_{3}}}{\alpha_{3}}\right) = 0,
\end{eqnarray}
\begin{eqnarray}
\label{18}
 \left(\frac{\dot{\alpha_{1}}}{\alpha_{1}} - \frac{\dot{\alpha_{3}}}{\alpha_{3}}\right) \frac{\dot{V}}{V} + \frac{d}{dt}\left(\frac{\dot{\alpha_{1}}}{\alpha_{1}} - \frac{\dot{\alpha_{3}}}{\alpha_{3}}\right)  = 0.
\end{eqnarray}
Integrating equations (14)-(16) and using Eqs. (9) and (10), the scale factors $\alpha_{1}(t),~ \alpha_{2} (t)$ and $\alpha_{3}(t)$ are able to be expressed clearly as
\begin{eqnarray}
\label{19}
\alpha_{1}(t) = V^{\frac{1}{3}},
\end{eqnarray}
\begin{eqnarray}
\label{20}
\alpha_{2}(t) = D V^{\frac{1}{3}} \exp{\left(X \int{\frac{dt}{V}}\right)}, 
\end{eqnarray}
\begin{eqnarray}
\label{21}
\alpha_{3}(t) = D^{-1} V^{\frac{1}{3}} \exp{\left(-X \int{\frac{dt}{V}}\right)} ,
\end{eqnarray}
where the integration constants are $X$ and $D$. The following gives the HDE density:
\begin{eqnarray}
\label{22}
\rho = 3(\delta H^{2} + \beta \dot{H}),
\end{eqnarray}
i.e $\rho = 3(\delta H^{2} + \beta \dot{H})$ with $M_{p}^{-2} = 8 \pi G = 1$ \cite{refN52a}. For the case of mutual interaction between DE and DM, the total ED $(\rho = \rho_{dm} + \rho)$ satisfies th continuity equation as
\begin{eqnarray}
\label{23}
\dot{\rho_{dm}} +\left (\frac{\dot{V}}{V}\right) \rho_{dm} = Q,
\end{eqnarray} 
\begin{eqnarray}
\label{24}
\dot{\rho} +\left (\frac{\dot{V}}{V}\right)(1+\omega) \rho = -Q,
\end{eqnarray} 
where $Q > 0$ indicates the strength of the interaction and $\omega = \frac{~~p}{\rho}$ is EoS parameter for HDE. The distinct conservation of DM and DE is implied by a vanishing $Q$. Given continuity equations, the energy density times a number with inverse time units-which might be the Hubble parameter $H$-must determine the interaction between DE and DM. The ED can take on any shape and can be any combination of dark matter and dark energy. Consequently, the phenomenological expressions of the interplay between DE and DM could be as follows:
\begin{eqnarray}
\label{25}
Q = 3 b^{2} H \rho_{dm} = b^{2} \frac{\dot{V}}{V} \rho_{dm},
\end{eqnarray}
where $b^{2}$ is coupling constant.
To circumvent the coincidence problem, Cai and Wang \cite{refN52b} have adopted the same relation for interacting DM and phantom DE.
We obtain the ED of dark matter using Eqs. (21) and (23).
\begin{eqnarray}
\label{26}
\rho_{dm} = \rho_{0} V^{b^{2}-1},
\end{eqnarray}
where the real constant of integration is $\rho_{0} > 0$.
Equations (23) and (24) allow us to obtain the interaction term $Q$ as
\begin{eqnarray}
\label{27}
 Q = 3 \rho_{0} b^{2} H V^{b^{2}-1}.
\end{eqnarray}
\section{Cosmological solutions }
Equations (4) through (8) can be solved by assuming the average scale factor (SF), which evolves with time evolution so that \cite{refaa53}:
\begin{equation}
\label{28}
a(t)=[\sinh(\alpha t)]^\frac{1}{n}.
\end{equation}
The main reason for employing this hyperbolic scale factor ansatz is its alignment with observations, and it has been utilized in building various cosmological models in different gravity theories \cite{pr,senta,sent,sent11,sz,sen,ent2,br1,n1,n2,cycl,n3}. Since cosmic transit from a decelerating phase to an accelerating phase is confirmed by recent observations \cite{refaa2}, we can utilize a form of the scale factor that results in a positive-to-negative sign flipping of the DP ($q$). The hyperbolic scale factor has been used in several cosmological contexts such as Bianchi cosmology \cite{pr}, universal extra dimensions \cite{n1}, $\kappa(R,T)$ gravity \cite{n2} and the cosmological models in Chern-Simons gravity \cite{sent,sent11}. Motivated by its consistency with observations, Sen and Sethi \cite{sen} used this hyperbolic scale factor in constructing an interesting quintessence model.  On getting motivated from increasing evidence for the need of a geometry that resembles Bianchi morphology to explain the observed anisotropy in the WMAP data, Pradhan et al.\cite{A1,A2} discussed the Bianchi type-$VI0$ and Bianchi type-I universes respectively with a time-dependent EoS parameter in general relativity and derived the hyperbolic scale factor $a(t) = sinh(\alpha T)$, where $T = t+ k$ and $\alpha$ is an arbitrary constant and discussed the dark energy model. Chawla et al. \cite{A3}  investigated a spatially homogenous and anisotropic Bianchi-I space-time with perfect fluid distribution, providing a cosmological model of massive strings with variable gravitational constant G and decaying vacuum energy density $\Lambda$ in general relativity. Einstein's field equations were solved in \cite{A3} by using a time-dependent deceleration parameter, yielding a scale factor $a = [Sinh(\alpha t)]^{1/n}$, where n is a positive constant. A unified cosmic description up to the late-time future has been presented in  \cite{sz} through utilizing a hyperbolic scale factor. Such hyperbolic ansatz also leads to  a desired behavior of the jerk parameters $j$. \\

From Eqs. (10) and (26), the volume scale factor
\begin{eqnarray}
\label{29}
V = \big[\sinh(\alpha t)\big]^\frac{3}{n}.
\end{eqnarray}
Using (27) in equations (17) - (19), we obtain the exact solutions of the SF as:
\begin{eqnarray}
\label{30}
\alpha_{1}(t)= [\sinh(\alpha t)]^\frac{1}{n},
\end{eqnarray}
\begin{eqnarray}
\label{31}
\alpha_{2}(t) = D [\sinh(\alpha t)]^{\frac{1}{n}}~ \exp{\left (X\int{[\sinh(\alpha t)]^{\frac{-3}{n}}} dt\right) },
\end{eqnarray}
\begin{eqnarray}
\label{32}
\alpha_{3} (t) = \frac{[\sinh(\alpha t)]^{\frac{1}{n}}~ \exp{\left (-X\int{[\sinh(\alpha t)]^{\frac{-3}{n}}} dt\right) }}{D}.
\end{eqnarray}
Using (27) in Equations.(24) and (25), we get
\begin{eqnarray}
\label{33}
\rho_{dm}  = \rho_{0}~[\sinh(\alpha t)]^{\frac{3}{n} (b^{2}-1)},
\end{eqnarray}

\begin{eqnarray}
\label{34}
Q = \frac{3\rho_{0} \alpha b^{2}}{n}~ \cosh(\alpha t)~ [\sinh(\alpha t)]^{-1+\frac{3}{n}(1+b^{2})}.
\end{eqnarray}
Using Equations. (28) - (31) in (4), we get the ED of HDE as
\begin{eqnarray}
\label{35}
\rho = \frac{3~\alpha^{2} \Big[\delta[\cosh(\alpha t)]^{2} - n \beta  \Big]}{n^{2} \Big[[\cosh (\alpha t)]^{2}-1\Big]}.
\end{eqnarray}
\begin{figure}[htbp]
	\centering
(a)	\includegraphics[width=7.5cm,height=9cm,angle=0]{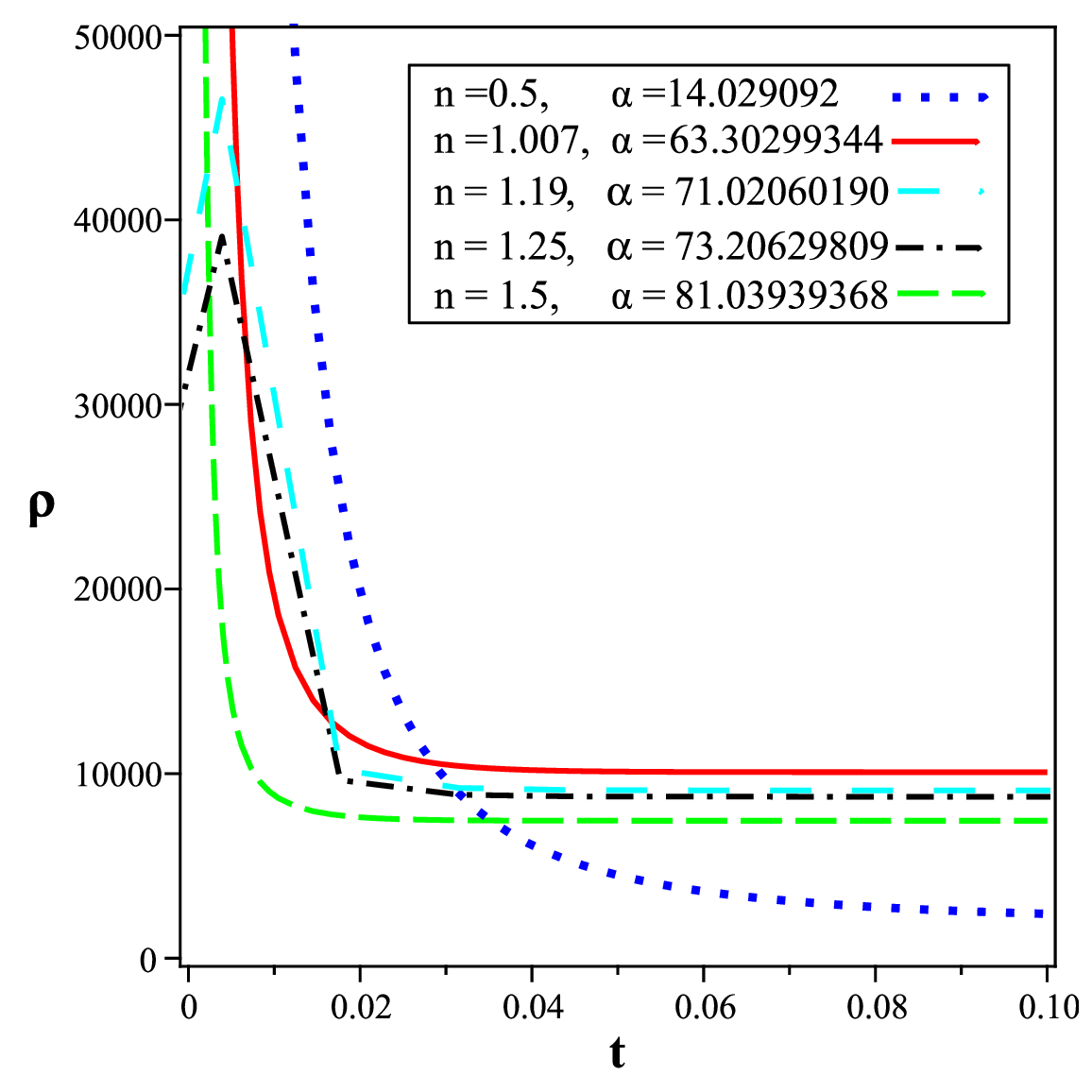}
(b)	\includegraphics[width=7.5cm,height=9cm,angle=0]{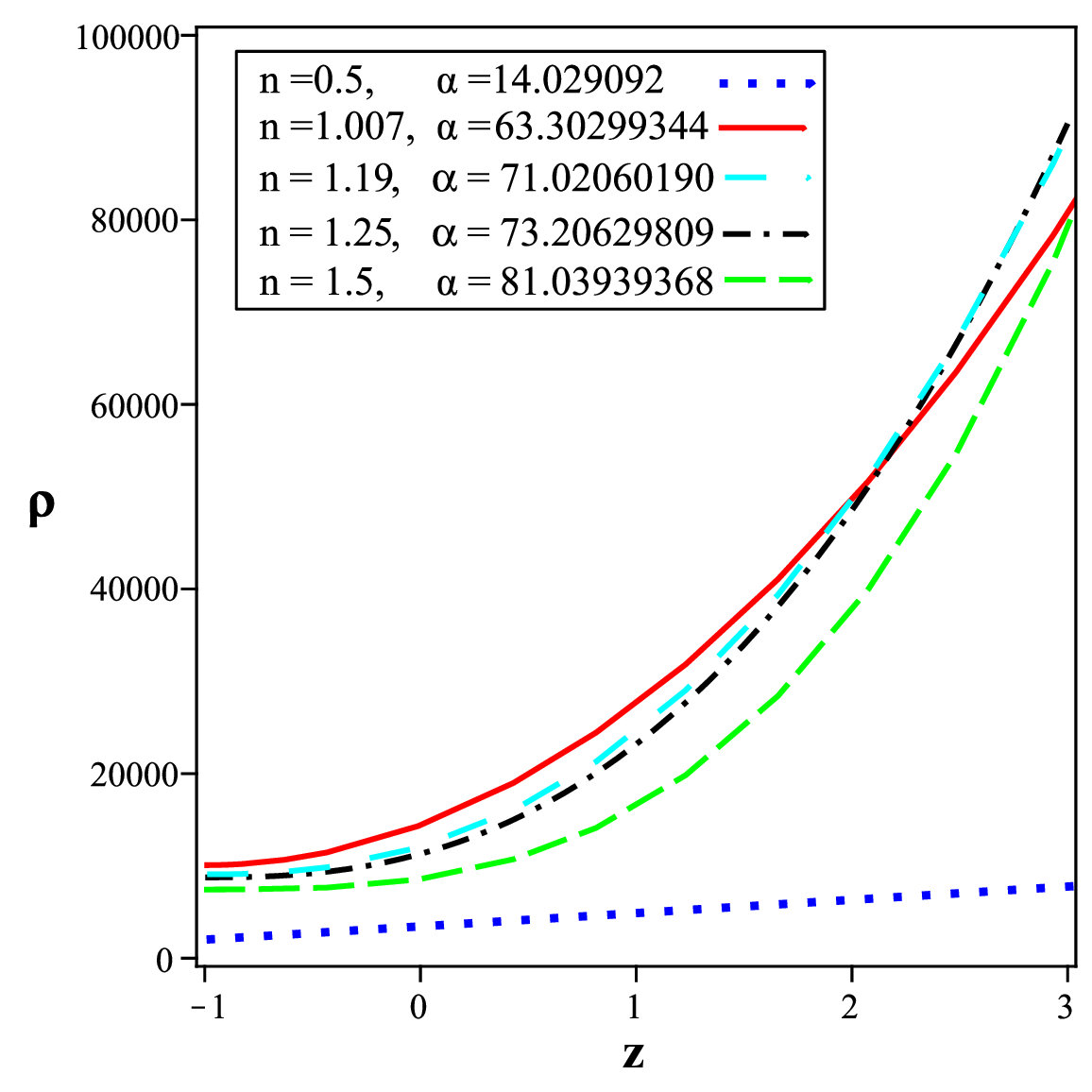}
	\caption{Plots of the ED of HDE  with (a) cosmic time $t$ and (b) redshift $z$ respectively for n= 0.5, 1.007, 1.19, 1.25, and 1.5.} 
\end{figure}\\
The graphs of energy density versus time ($t$) and redshift ($z$) are displayed separately in Figures 1(a) and 1(b). As we can see, the energy density is positive and a declining function that converges to zero as $t \to \infty$. Using Eqs. (27), (31) and (33) in Eqs. (4) - (8), we get the pressure of HDE as 
\begin{eqnarray}
\label{36}
~~p &=& \frac{1}{n^{2} \Big[[\cosh (\alpha t)]^{2}-1 \Big]} \Bigg[-n^{2} b^{2} \rho_{0}~\Big[[\cosh(\alpha t)]^{2} -1\Big] ~[\sinh(\alpha t)]^{\frac{3}{n} (b^{2}-1)} \nonumber \\ 
&-&
2 \alpha^{2}\left(\frac{3}{2} ~\delta ~[\cosh(\alpha t)]^{2} + n \left(\frac{-3}{2}\beta - \delta + n\beta \right)\right)\Bigg]
\end{eqnarray}\\
\begin{figure}[htbp]
	\centering
(a)	\includegraphics[width=7.5cm,height=9cm,angle=0]{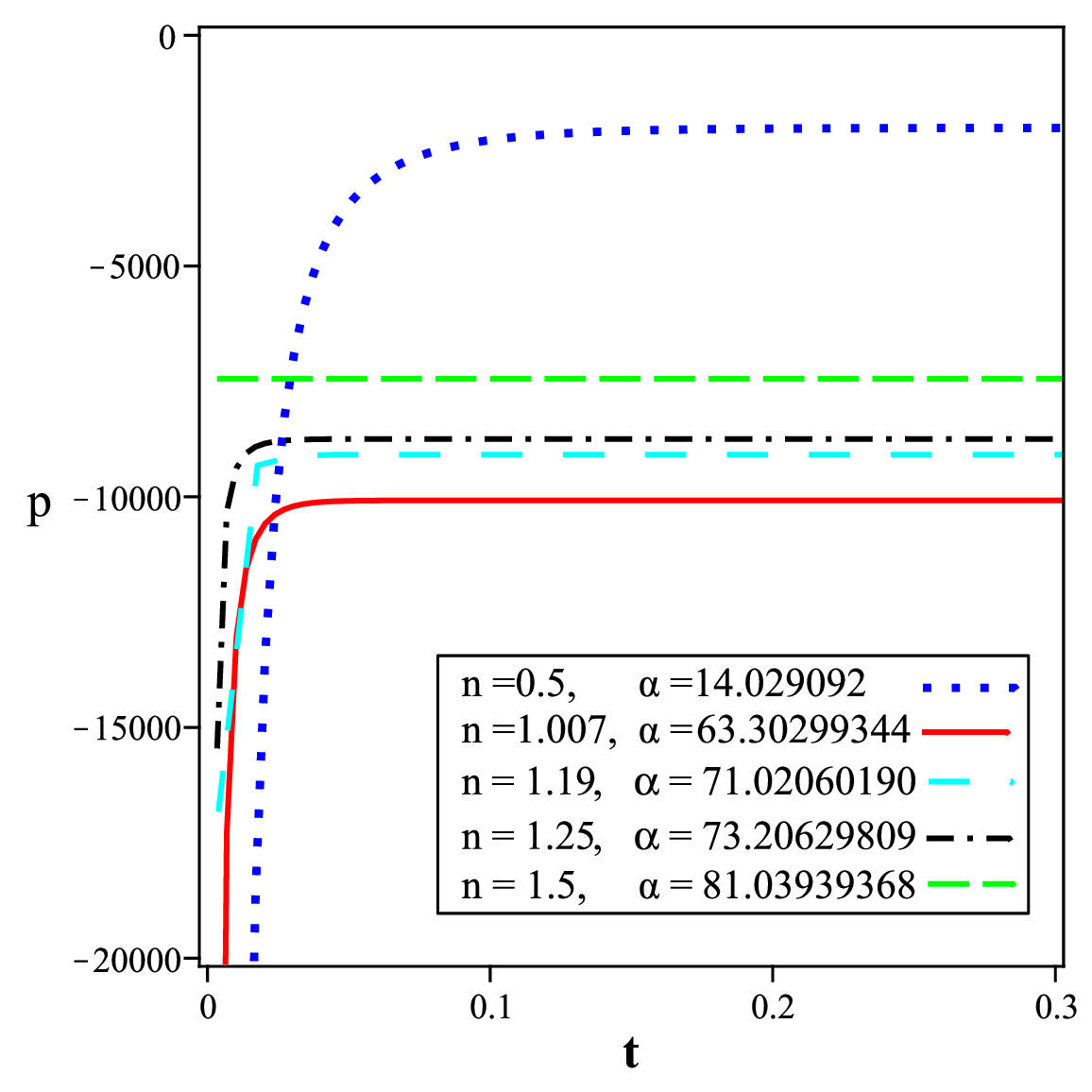}
(b)	\includegraphics[width=7.5cm,height=9cm,angle=0]{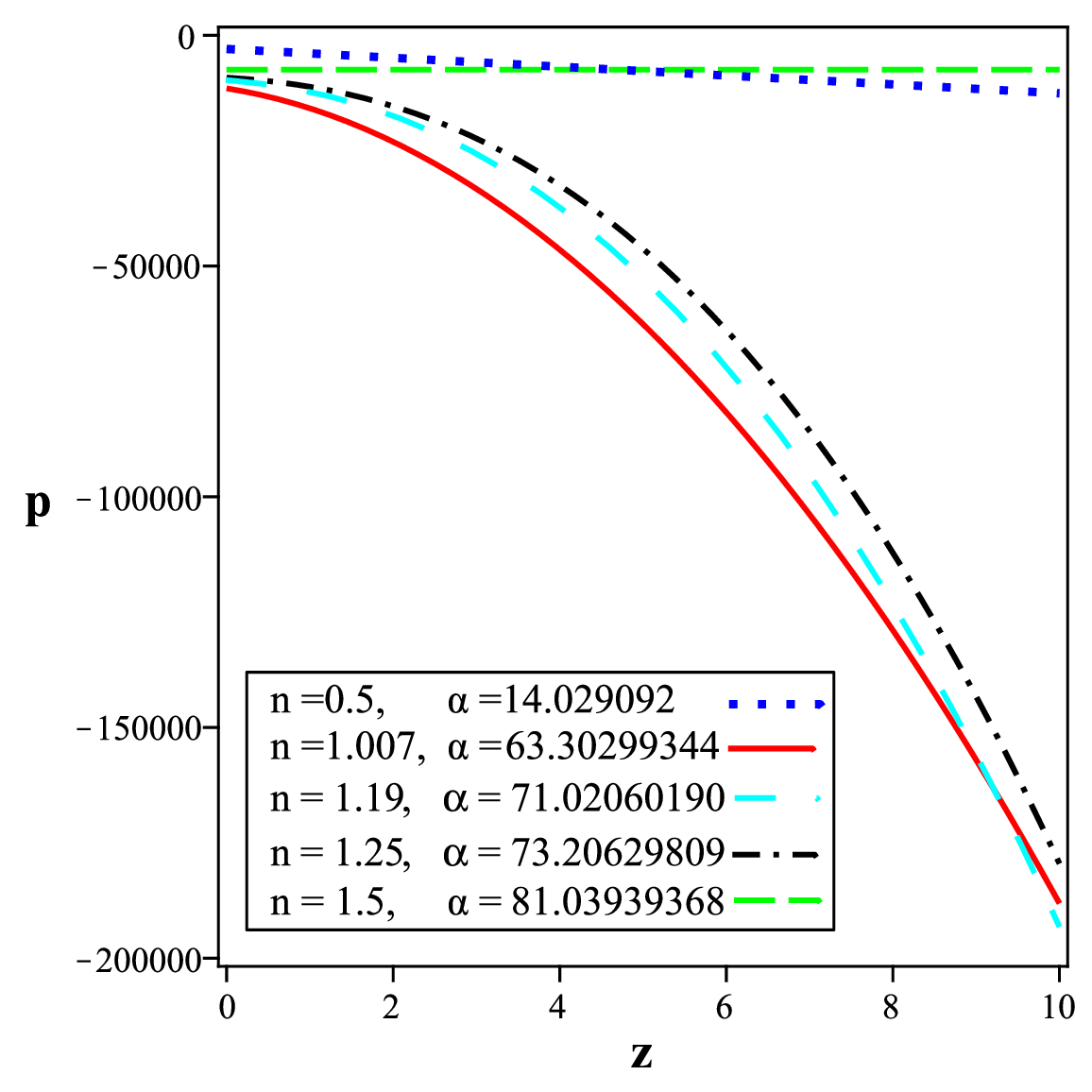}
	\caption {Plots of the HDE pressure for n = 0.5, 1.007, 1.19, 1.25, and 1.5 with respect to (a) cosmic time $t$ and (b) redshift $z$ respectively. } 
\end{figure}\\
The pressure fluctuation with respect to time (t) and redshift (z) for n = 0.5, 1.007, 1.19, 1.25, and 1.5 is explained in Figure 2(a) and (b). According to the numbers, pressure increases with time. It reaches a small negative number close to zero after starting with a large negative value. Following the discovery of the universe's extended acceleration, the prevailing consensus is that "dark energy"-a negative-pressure energy-matter-is responsible for this cosmic acceleration. The EoS parameter of HDE can be found as
\begin{eqnarray}
\label{37}
\omega &=& \frac{1}{3 n^{2}\Big[-\delta [\cosh(\alpha t)]^{2} + n \beta \Big]}  
\Bigg[n^{2} b^{2} \rho_{0}~\Big[[\cosh(\alpha t)]^{2} -1\Big] ~[\sinh(\alpha t)]^{\frac{3}{n} (b^{2}-1)} \nonumber \\ 
&+& 
2 \alpha^{2}\left(\frac{3}{2} ~\delta ~[\cosh(\alpha t)]^{2} + n \left(\frac{-3}{2}\beta - \delta + n\beta \right)\right)\Bigg]
\end{eqnarray}
\begin{figure}[htbp]
	\centering
(a)	\includegraphics[width=7.5cm,height=9cm,angle=0]{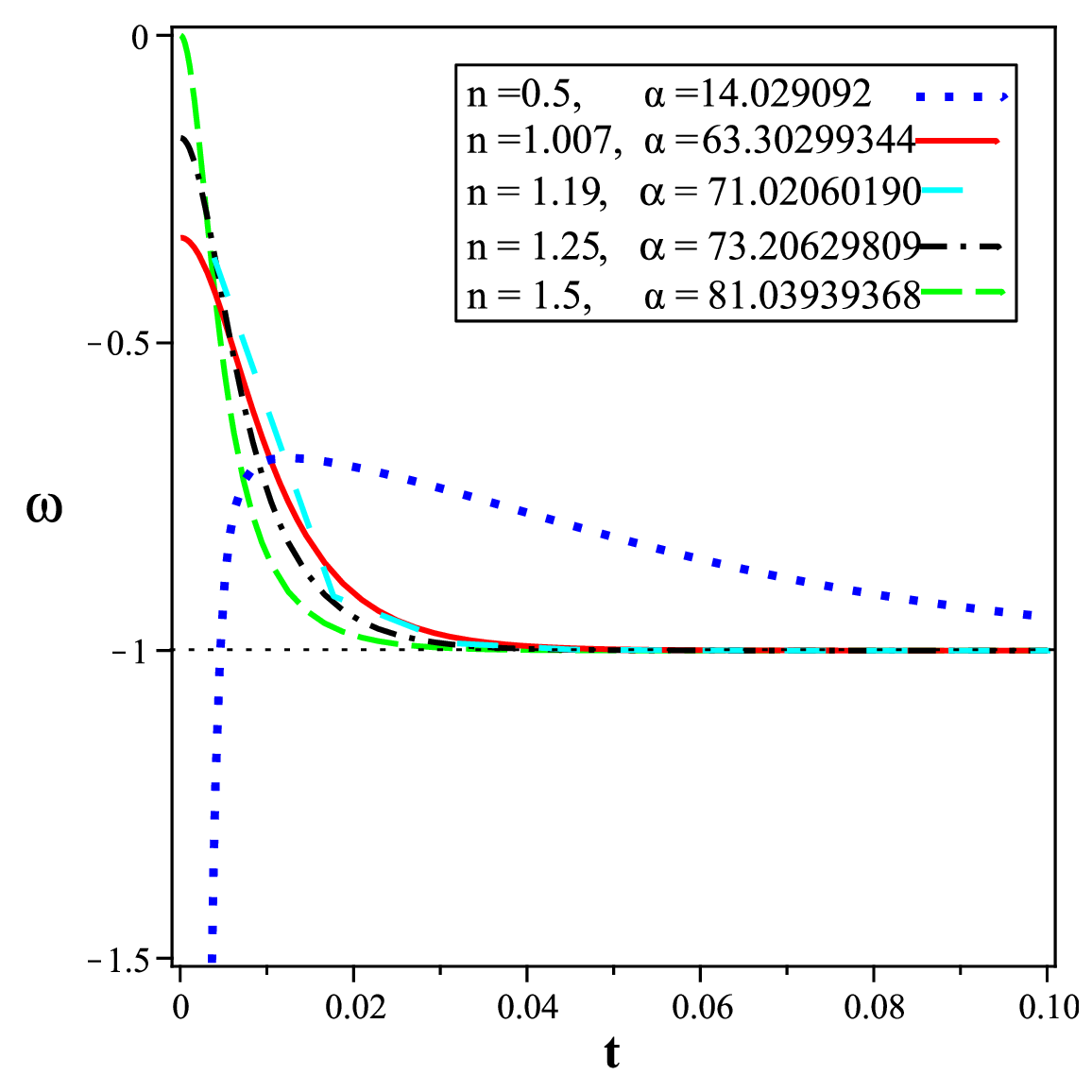}
(b)	\includegraphics[width=7.5cm,height=9cm,angle=0]{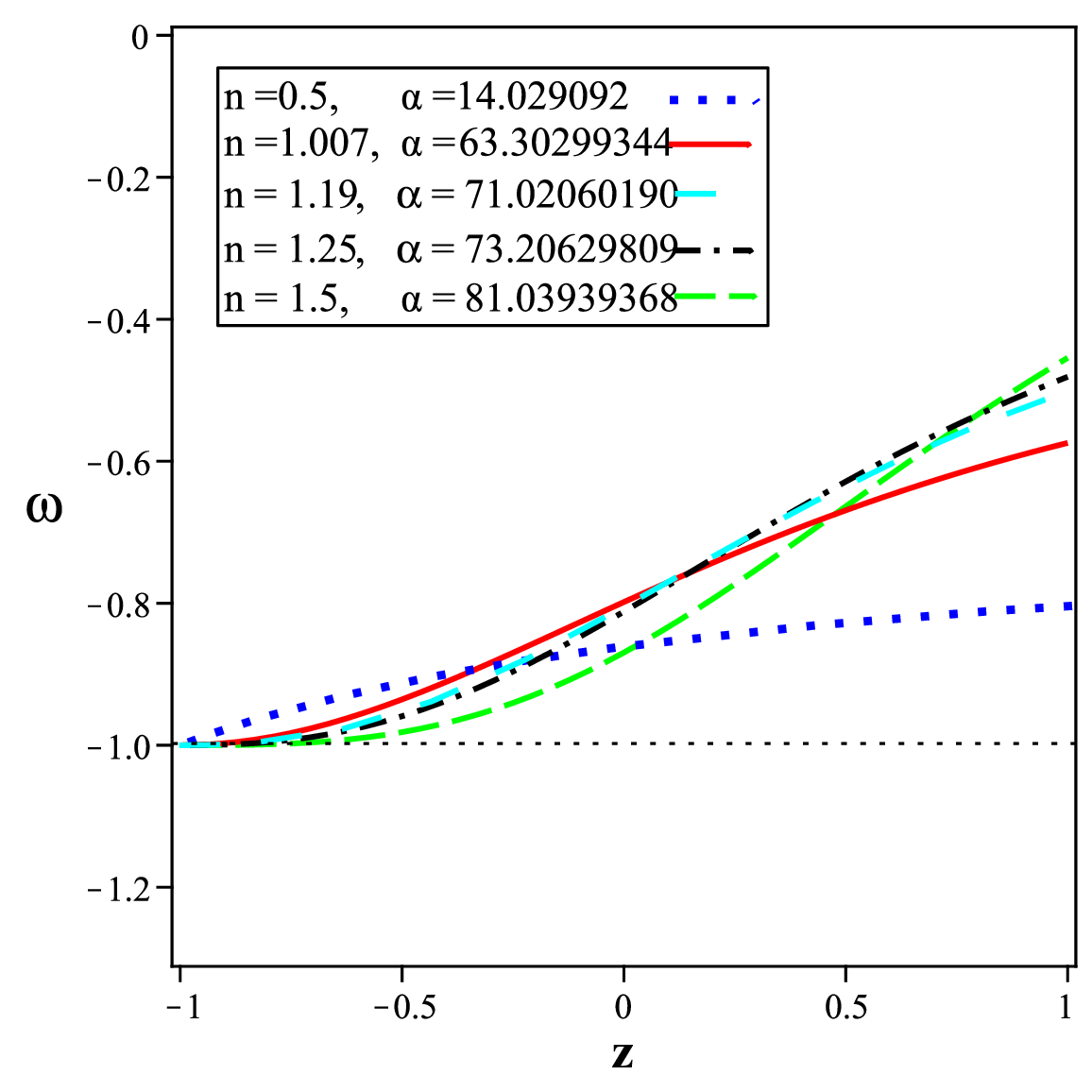}
	\caption{Plot of the EoS parameter of HDE with cosmic time $t$ in fig(a) and with redshift $z$ in fig(b) for n= 0.5, 1.007, 1.19, 1.25, and 1.5 respectively} 
\end{figure}
Fig. 3(a) and Fig. 3(b) show the variation of the EoS parameter with cosmic time (t) and redshift (z). Both figures show that the EoS parameter remains in quintessence epoch ($\omega > -1$) and tends to the cosmological constant ($\omega = -1$) at the future for n= 0.5, 1.007, 1.19, 1.25, and 1.5. It is noteworthy that the EoS parameter shows quintessence-like nature for the diverse estimations of $n$.  Astrophysical observations and the analysis of HDE inflation with Hubble's cut-off, as reported by \cite{refaa55,refaa56}, confirm the fact that the EoS parameter in our framework can transcend the phantom divided line.\\

Using equations (28) - (32) in equations (11) - (15). We find the mean Hubble parameter $H$, deceleration parameter $q$, mean anistropy parameter of expansion $(\Delta)$,  the expansion scalar $\theta$  and the shear scalar $\sigma^{2}$ respectively as
\begin{eqnarray}
\label{38}
H = \frac{\alpha  \cosh(\alpha t)}{n \sinh(\alpha t)}
\end{eqnarray}
\begin{eqnarray}
\label{39}
q = \frac{-\big[\cosh(\alpha t)\big]^{2} + n}{ \big[\cosh(\alpha t)\big]^{2}}.
\end{eqnarray}

We have acquired a relation among the parameters $n$ and $\alpha$ as $\alpha = \frac{1}{12.36} \tanh^{-1} \Big[\Big(1-\frac{0.48}{n}\Big)^{\frac{1}{2}}\Big]$
from the expression of DP. In the current work, we have considered the observational data based on the data combination with OHD and JLA \cite{refaa55}, where $q_{0} = -0.52$ and $H_{0}=69.2$. Combining Observational Hubble Data (OHD) and the Joint Light-curve Analysis (JLA) dataset provides more robust restrictions on cosmological parameters, notably those linked to dark energy and the universe's expansion rate. The OHD dataset, mostly derived from cosmic chronometers, gives direct Hubble parameter values, and the JLA dataset, based on Type Ia supernovae, provides information about luminosity distances and expansion history.  Combining these datasets helps to decrease the systematic errors inherent in each individual dataset and can lead to more precise estimates of parameters like the Hubble constant ($H_0$) and the deceleration parameter ($q_0$) (See Refs. \cite{refA4,refA5,refA6}).   \\


\begin{table}[h!]
	\begin{center}
		\caption{The behavior of Deceleration parameter for different values of $\alpha$ and $n$:}
		\label{tab:table1}
		\begin{tabular}{c|c|c|c|c} 
			\textbf{n} & \textbf{$\alpha$} & \textbf{q} & \textbf{Cosmic phase} & \textbf{z} \\
			\hline
			0.5 & 14.02909274 & ``(-ve) & ``Acceleration &  - \\
			0.75 & 47.96578492&  (-ve) & Acceleration & -\\
			1 & 62.968828 &  (-ve) & Acceleration & -\\
			1.007 & 63.30299344 & (+ve) to (-ve) & Deceleration to Acceleration & -\\
			1.02 & 63.91471456 & (+ve) to (-ve) & Deceleration to Acceleration & -\\
			1.07 & 66.16708343 & (+ve) to (-ve) & Deceleration to Acceleration & 2.3\\
			1.085 & 66.81393420 & (+ve) to (-ve) & Deceleration to acceleration & 2\\
			1.09 & 67.02677747 & (+ve) to (-ve) & Deceleration to Acceleration & 1.9\\
		    1.11 & 67.86476551 & (+ve) to (-ve) & Deceleration to Acceleration & 1.7\\
			1.2 & 71.39492255 & (+ve) to (-ve) & Deceleration to Acceleration & 0.97\\
			1.25 & 73.20629809 & (+ve) to (-ve) & Deceleration to Acceleration & 0.7\\
			1.27 & 73.90435295 &  (+ve) to (-ve) & Deceleration to Acceleration & 0.66\\
			1.3 & 74.92501503 & (+ve) to (-ve) & Deceleration to Acceleration & 0.59\\
			1.34 & 76.23953068 & (+ve) to (-ve) & Deceleration to Acceleration & 0.50\\
			1.42 & 78.72440767 & (+ve) to (-ve) & Deceleration to Acceleration & 0.36\\
			1.45 & 79.61114498 & (+ve) to (-ve) & Deceleration to Acceleration  & 0.33\\
			1.5 & 81.03939368 & (+ve) to (-ve) & Deceleration to Acceleration & 0.25\\
			1.75 & 87.39949592 & (+ve) to (-ve) & Deceleration to Acceleration  & 0.09\\
			2 & 92.758996 &  (+ve) to (-ve) & Deceleration to Acceleration & 0\\
			4 & 119.150718 & (+ve) to (-ve)" & Deceleration to Acceleration" & -0.12\\
		\end{tabular}
	\end{center}
\end{table}

\begin{figure}[htbp]
	\centering
(a)	\includegraphics[width=7.5cm,height=9cm,angle=0]{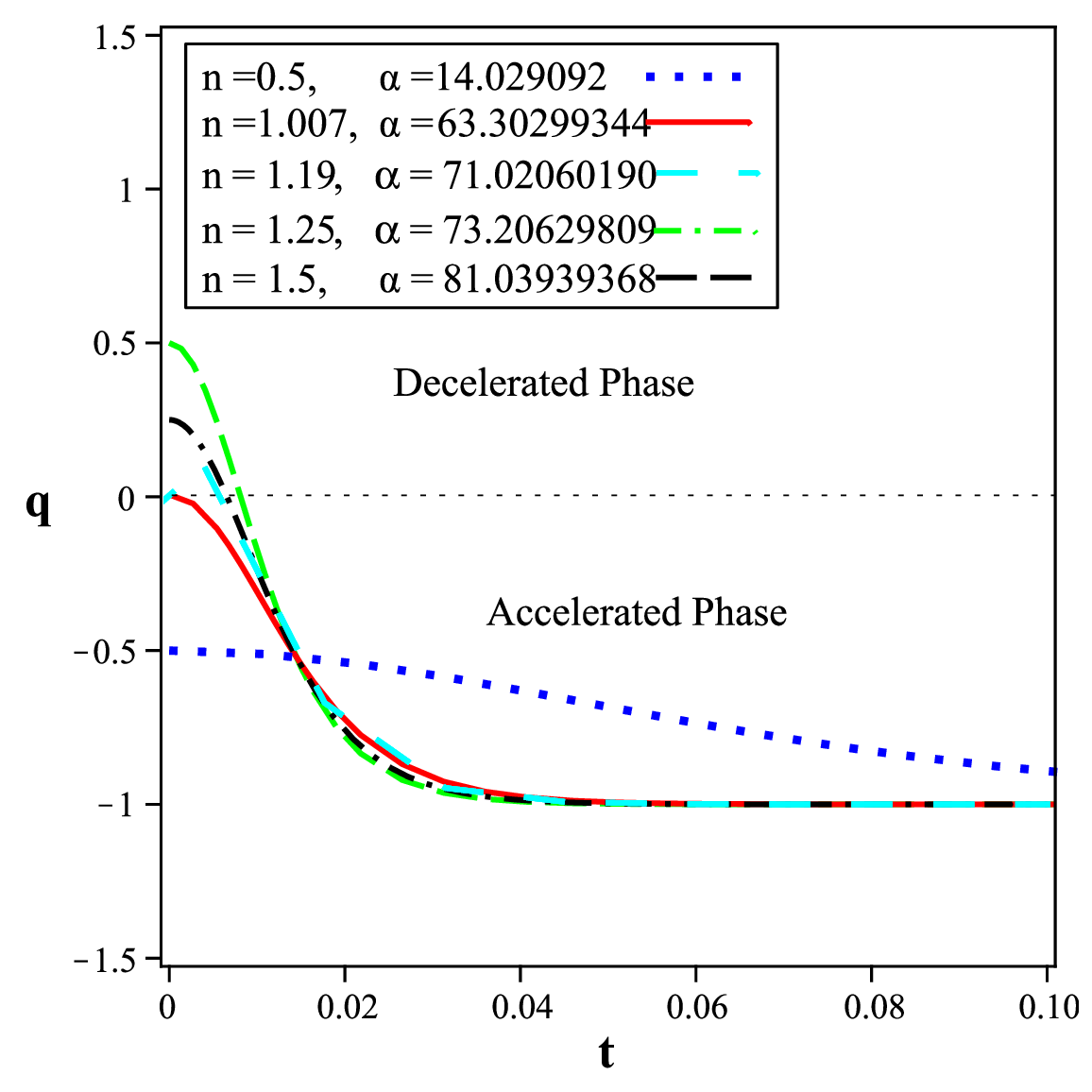}
(b)	\includegraphics[width=7.5cm,height=9cm,angle=0]{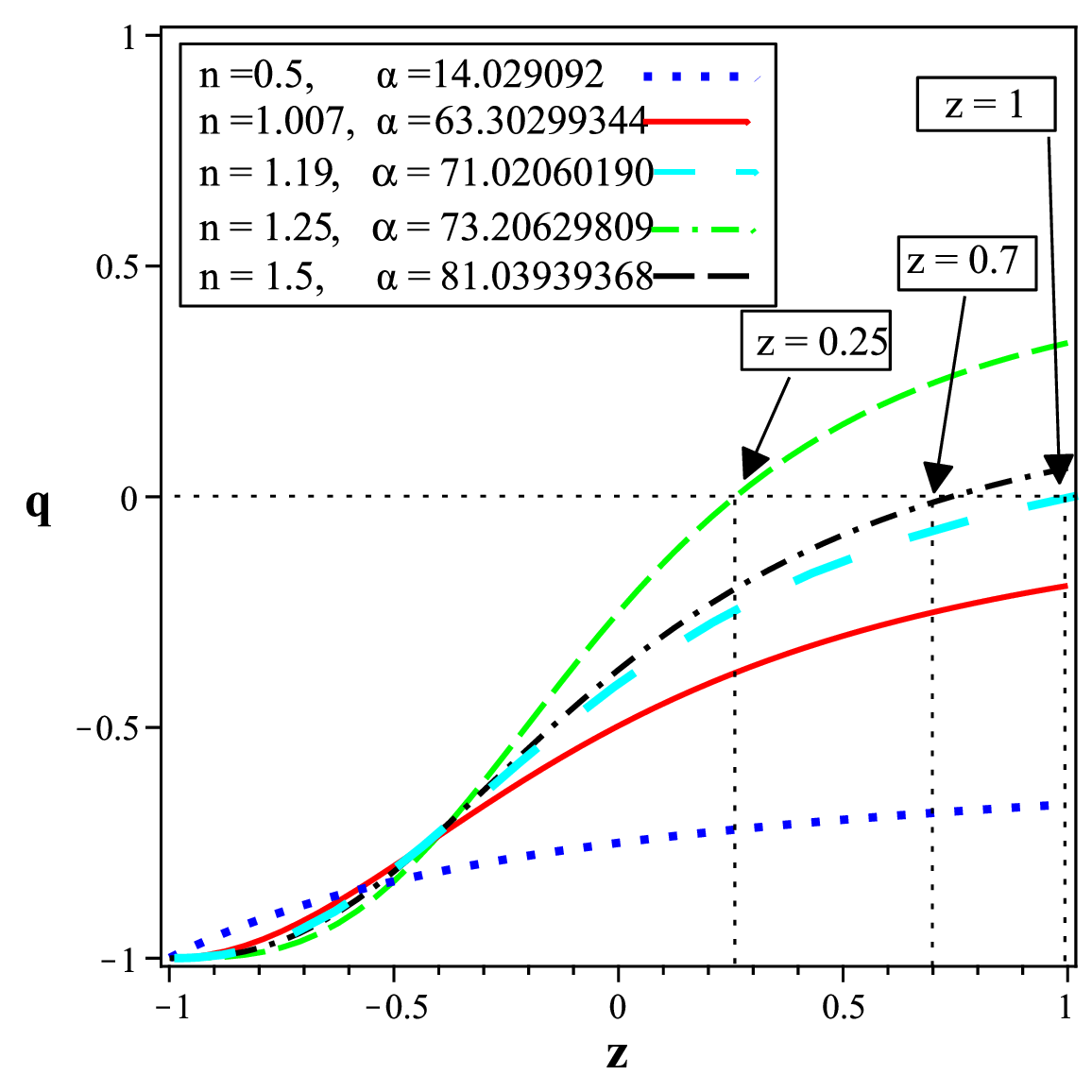}
	\caption{Plot of deceleration parameter $q$ with cosmic time (t) in fig(a) and redshift (z) in fig(b) respectively for n= 0.5, 1.007, 1.19, 1.25, and 1.5} 
\end{figure}

The evolutionary nature of the deceleration parameter (DP) is graphed for cosmic time (t) and redshift (z) in Fig. 4(a) and (b), respectively, for the values of $n = 0.5, 1.007, 1.19, 1.25$, and $1.5$. Based on the value for the DP (q), the accelerating phase happens for $q < 0$ and the decelerating phase for $q > 0$. observations shows that the value of DP lies in the range of $-1 \leq q < 0$. One can mark from the figures that the DP transits from an early decelerating to the recent accelerating era for diverse values of n. 
\begin{eqnarray}
\label{40}
\Delta = \frac{2 X^{2} n^{2}\big[\sinh(\alpha t)\big]^{\frac{2(n-3)}{n}}}{3 \alpha^{2} \big[\cosh(\alpha t)\big]^2} 
\end{eqnarray}
\begin{eqnarray}
\label{41}
\theta =  \frac{3 \alpha  \cosh(\alpha t)}{n \sinh(\alpha t)}
\end{eqnarray}
\begin{eqnarray}
\label{42}
\sigma^{2} = \frac{X^{2} \big[\sinh(\alpha t)\big]^{\frac{2(n-3)}{n}}}{\big[\cosh(\alpha t)\big]^{2} -1}
\end{eqnarray}
The coincidence parameter $r = \frac{\rho_{dm}}{\rho}$  is the ratio of dark matter energy density to the dark energy density. This should stays constant or deviates very slowly near the current period. Though, the famous $\Lambda$CDM model is not consistent with this observation for the leading candidate of DE. Multiple researchers have been directed by this coincidence problem to view alternatives to $\Lambda$CDM that maintain its astonishing victories (CMB anisotropies, large-scale structure, Type Ia SNe) but sidesteps the above problems. For the large stage expansion of the Universe, matter and energy must mount each other over a largely long duration of time to avoid the coincidence problem. It is stated as:
\begin{eqnarray}
\label{43}
r = \frac{n^{2} \rho_{0}\big[\sinh(\alpha t)\big]^{\frac{3}{n}(b^{2}-1)} \Big[\big[\cosh(\alpha t)\big]^{2} -1\Big]}{3 \alpha^{2}  \Big[\delta\big[\cosh(\alpha t)\big]^{2} -n \beta\Big]}
\end{eqnarray}
\begin{figure}[htbp]
	\centering
(a)	\includegraphics[width=7.5cm,height=9cm,angle=0]{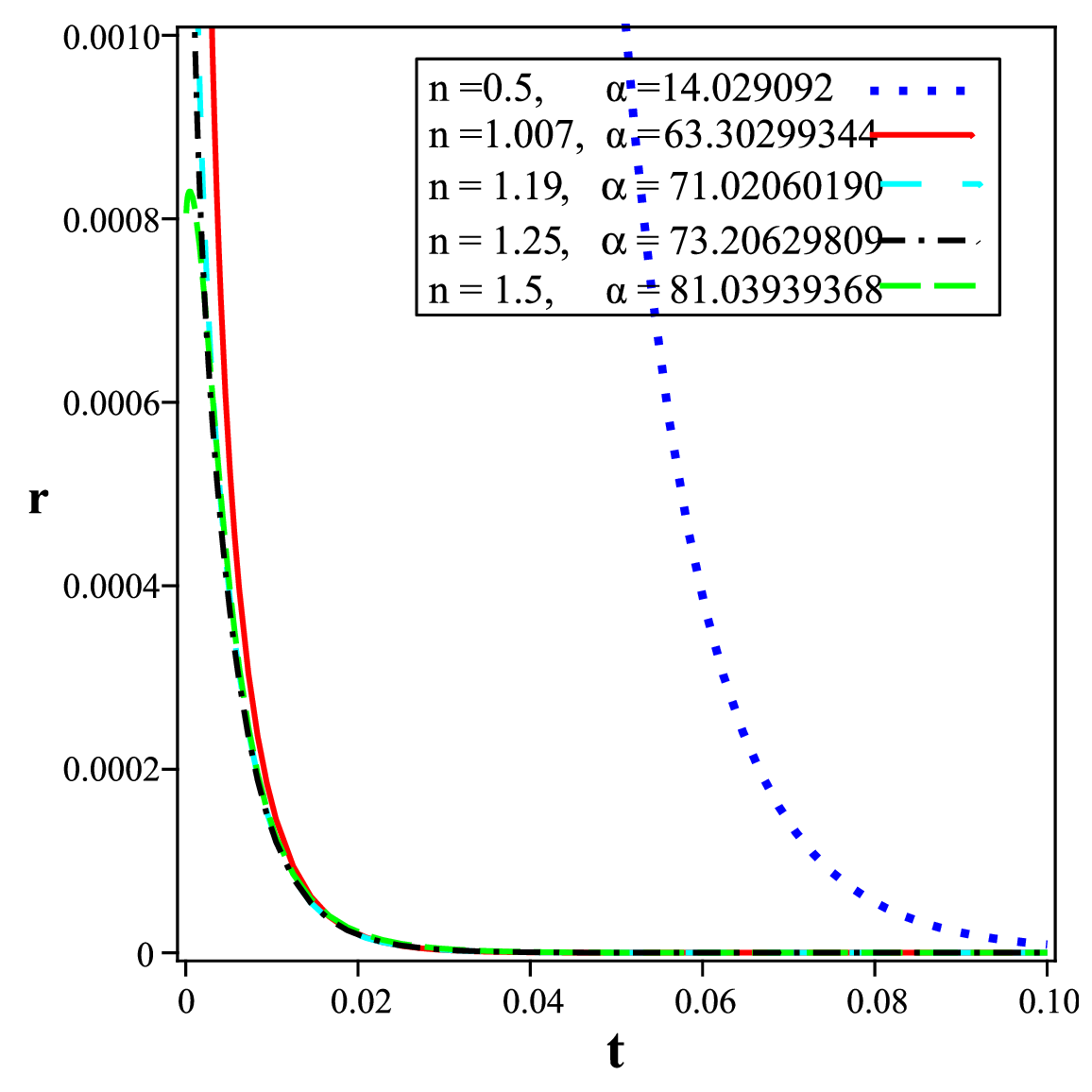}
(b)	\includegraphics[width=7.5cm,height=9cm,angle=0]{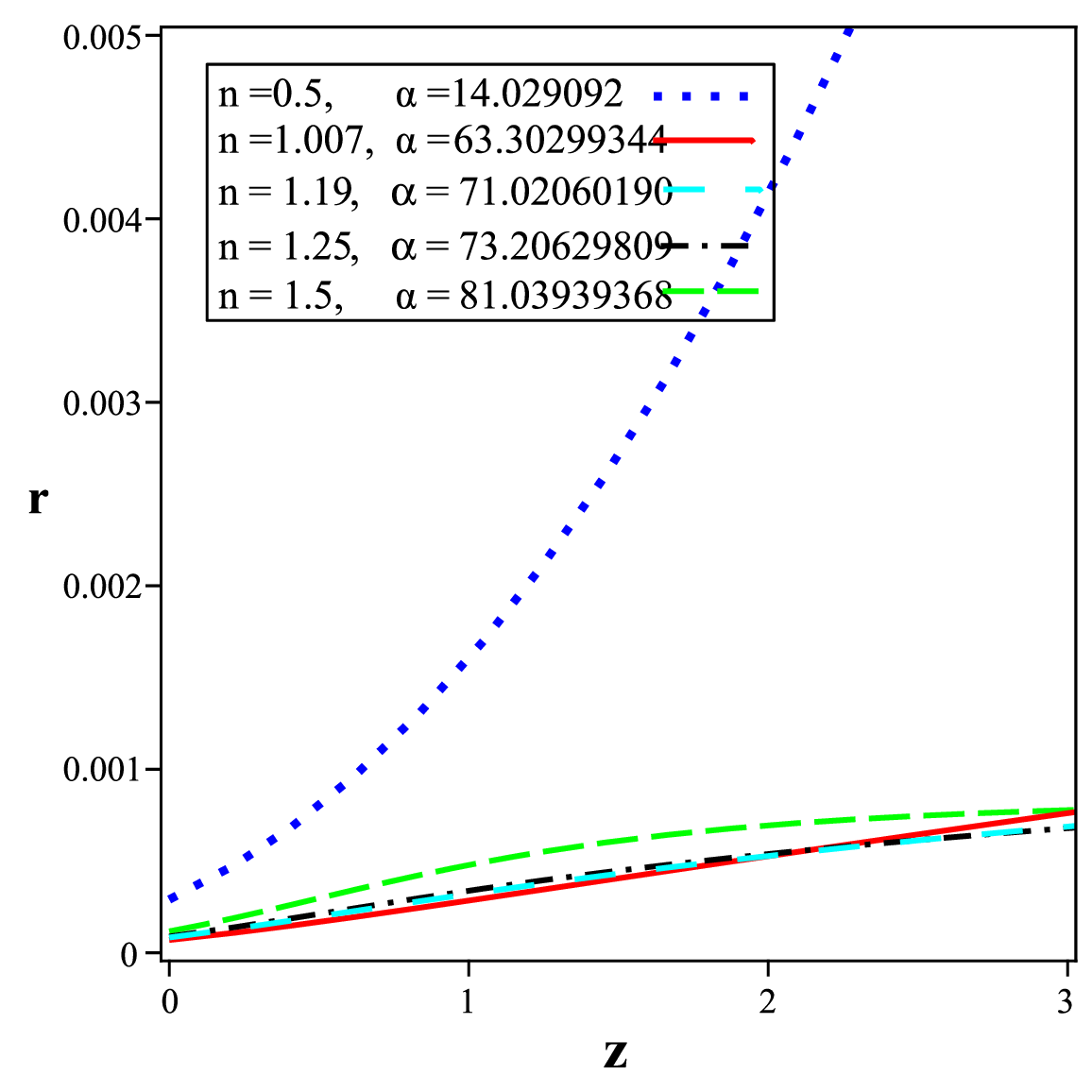}
	\caption{Plot of the coincidence parameter $r$ with cosmic time t in fig(a) and redshift z in fig(b) respectively for n= 0.5, 1.007, 1.19, 1.25, and 1.5} 
\end{figure}
The variation of the coincidence problem is shown in Fig. 5(a) and Fig. 5 (b) versus cosmic time (t) and redshift (z), respectively for n =  0.5, 1.007, 1.19, 1.25, and 1.5. We can see that the coincidence parameter (r) varies at a very early stage and then converges to a constant value at late-times. Therefore, an appropriate sort of interaction between HDE and DM can resemble the ratio of their densities likely to acquire a fixed value under the expansion and accordingly can assist in soothing the coincidence problem that arises in the $\Lambda$CDM model.

\section{Physical acceptability of the solutions}
\subsection{Speed of sound}
Causality implies that the sound speed ($v_{s}$) must be less than the speed of light (c). In gravitational units, the sound speed exists within the range $0 \leq v_{s} = \frac{dp}{d\rho} \leq 1$ and is given by
\begin{eqnarray}
\label{44}
v_{s} = \frac{dp}{d\rho}
\end{eqnarray}
where, 
\begin{eqnarray}
\label{45}
v_{s} = -\frac{1}{2} \frac{1}{\alpha^{2} (-\delta + \beta n)} \bigg[n b^{2} (\sinh(\alpha t))^\frac{-3}{n}~~ p~~ (\cosh^{2}(\alpha t) - 1)(b^{2} -1) (\sinh(\alpha t))^\frac{-3b^{2}}{n} \\ \nonumber
 - \frac{4}{3}\alpha^{2} \bigg[-\frac{3}{2} +n \bigg] (-\delta + \beta n) \bigg]
\end{eqnarray}

\begin{figure}[htbp]
	\centering
(a)	\includegraphics[width=7.5cm,height=9cm,angle=0]{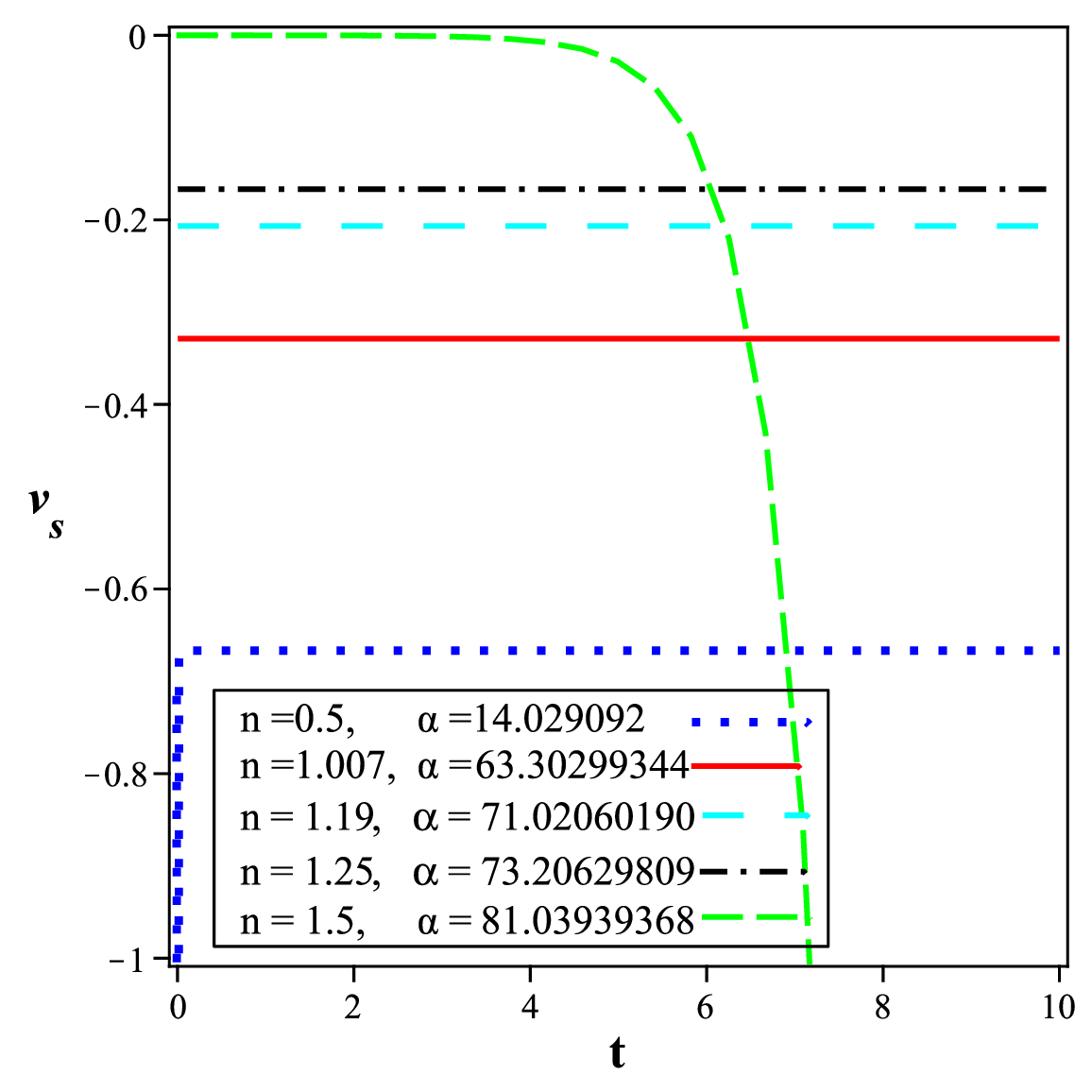}
(b)	\includegraphics[width=7.5cm,height=9cm,angle=0]{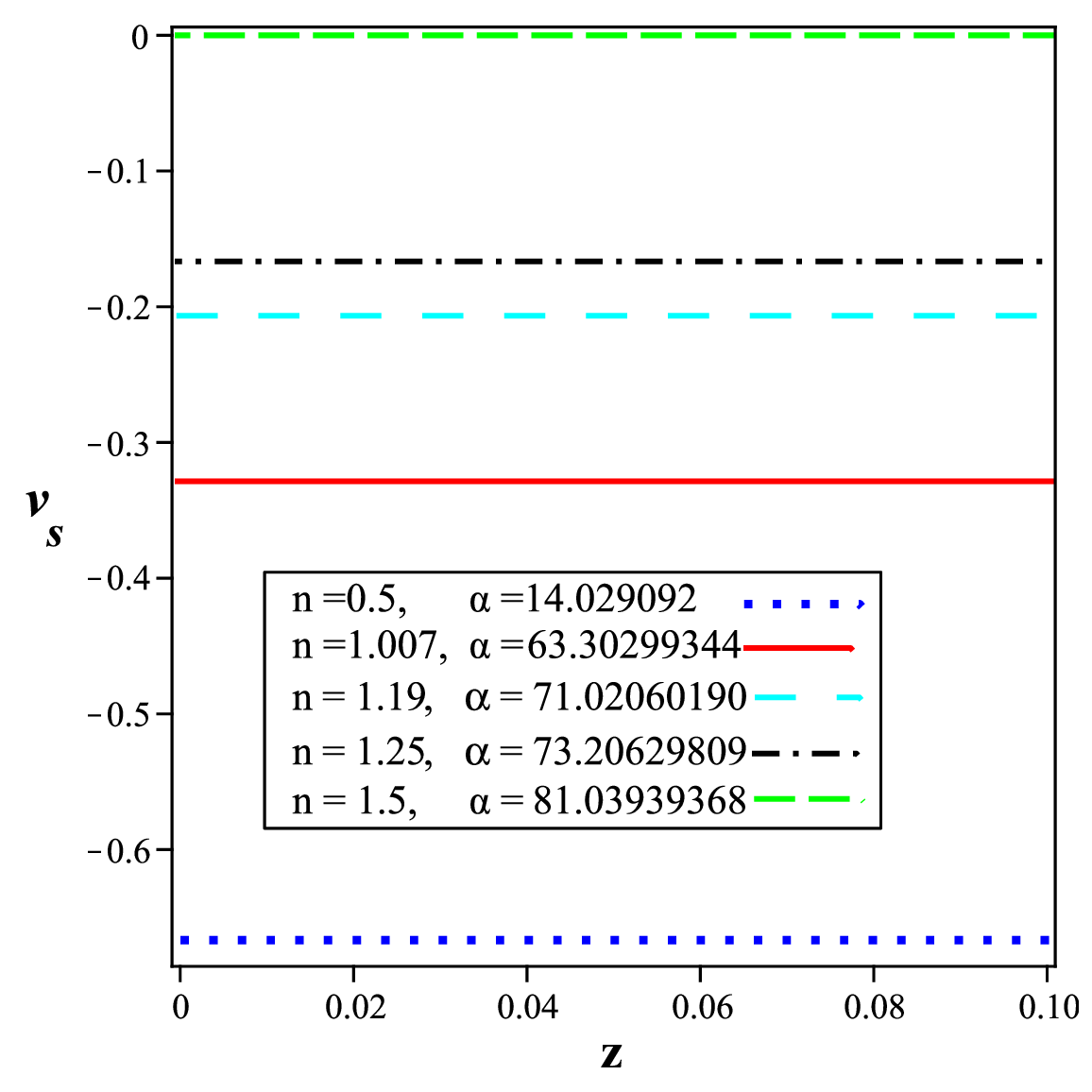}
	\caption{Plot of speed of sound $v_{s}$ with cosmic time (t) in fig(a) and redshift (z) in fig(b) respectively for n= 0.5, 1.007, 1.19, 1.25, 1.5} 
\end{figure}
Here we observe that $v_{s} < 1$. Figs. 6 (a) and 6 (b) show the plot of speed of sound with time (t) and redshift (z) respectively.
\subsection{Energy conditions}
The classical linear energy conditions \cite{ec11} provide another way to test the physical acceptability of the current model. The weak, null and dominant energy conditions are given as (i) ~~$\rho \geq 0$(WEC),~~~(ii) ~~$\rho+p \geq 0$(NEC),~~~(iii) ~~$\rho-p \geq 0$(DEC). The strong energy conditions (SEC) can be written as $\rho + 3p \geq 0$ and it implies the attractive nature of gravity. So, in general, this condition is not expected to be valid for an accelerating or inflationary universe \cite{ec3}.
\begin{figure}[htbp]
	\centering
(a)	\includegraphics[width=7.5cm,height=9cm,angle=0]{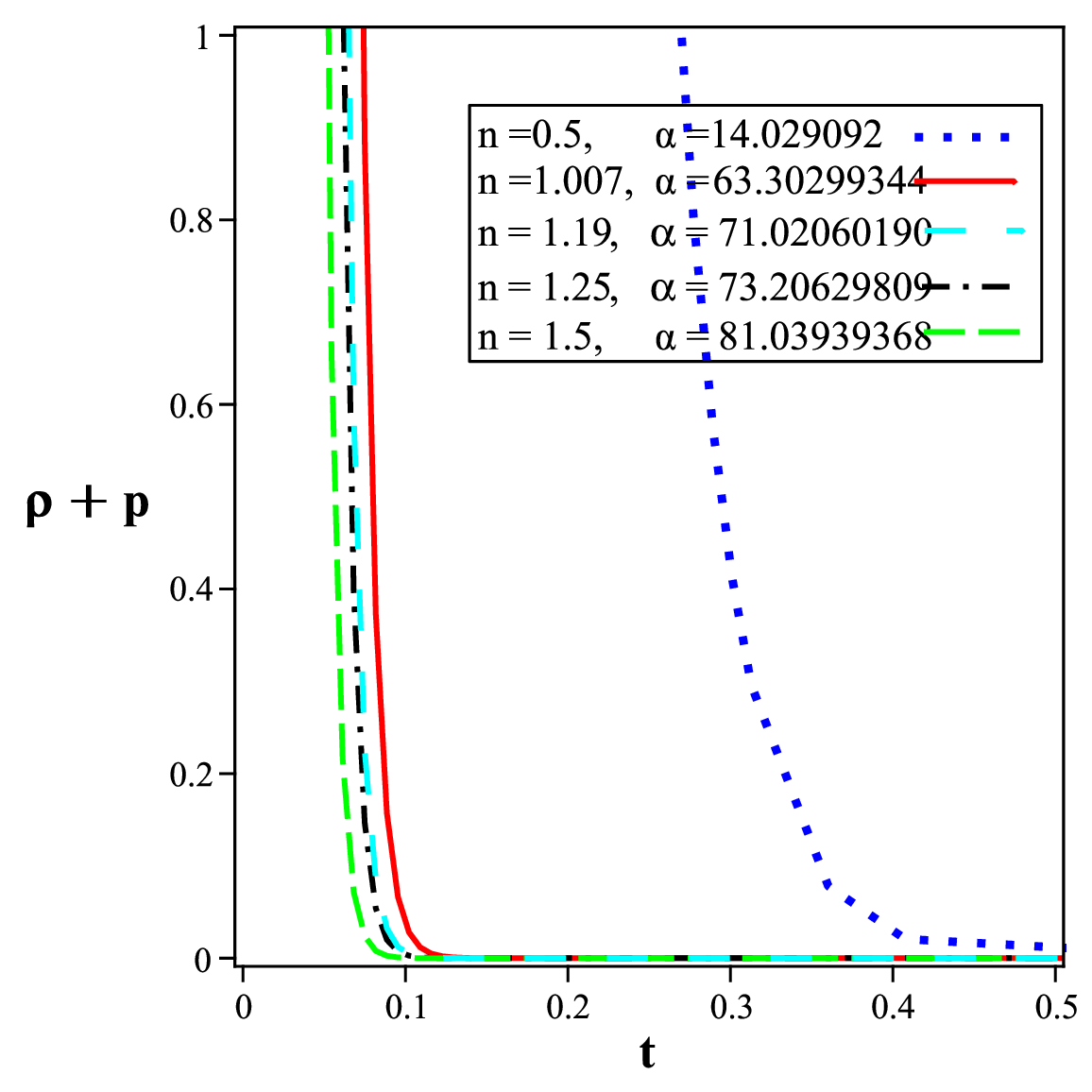}	(b)\includegraphics[width=7.5cm,height=9cm,angle=0]{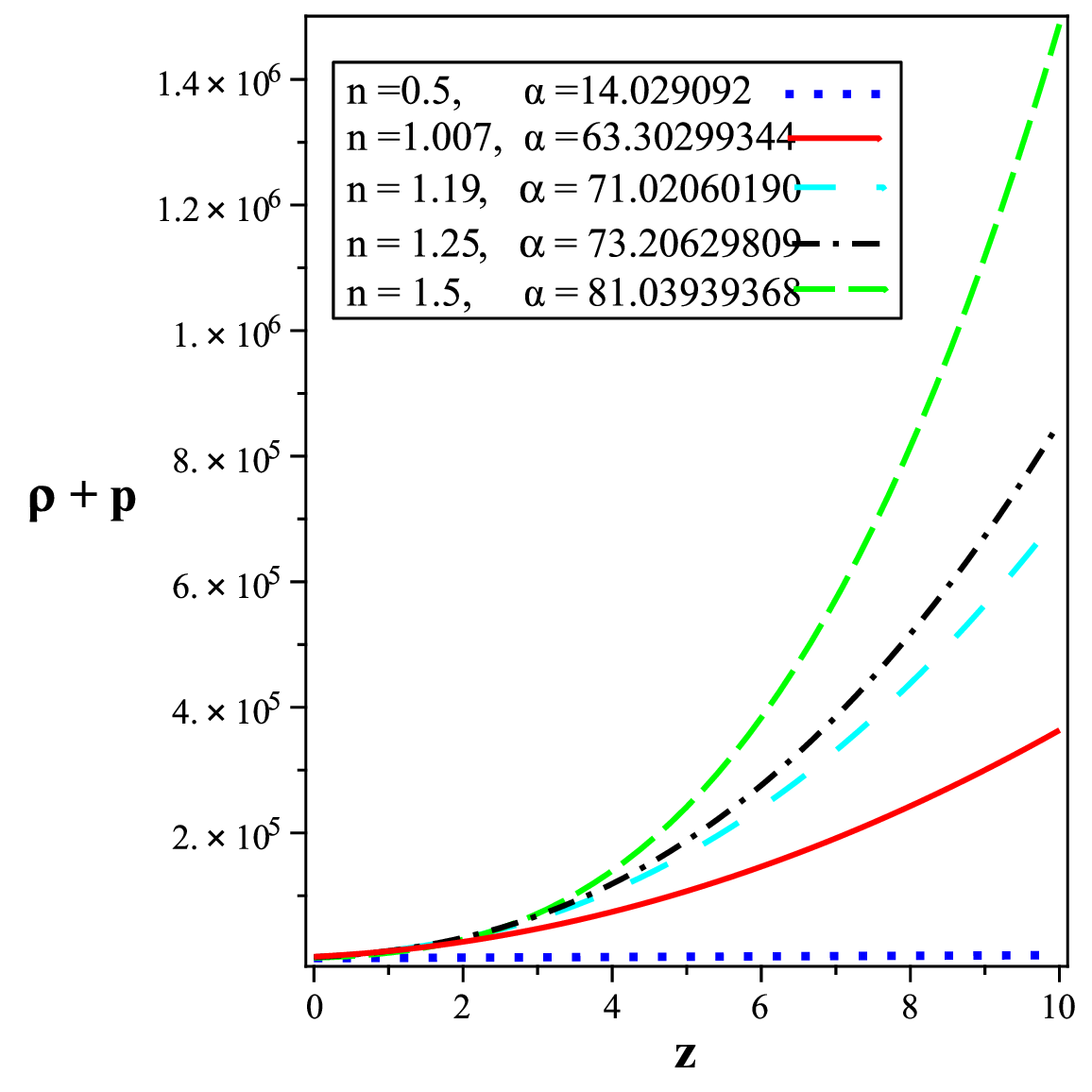}
	\caption{Plot of energy condition($\rho+p$) with cosmic time (t) in fig(a) and redshift (z) in fig(b) respectively for n= 0.5, 1.007, 1.19, 1.25, 1.5} 	
\end{figure}	
\begin{figure}[htbp]	
(a)	\includegraphics[width=7.5cm,height=9cm,angle=0]{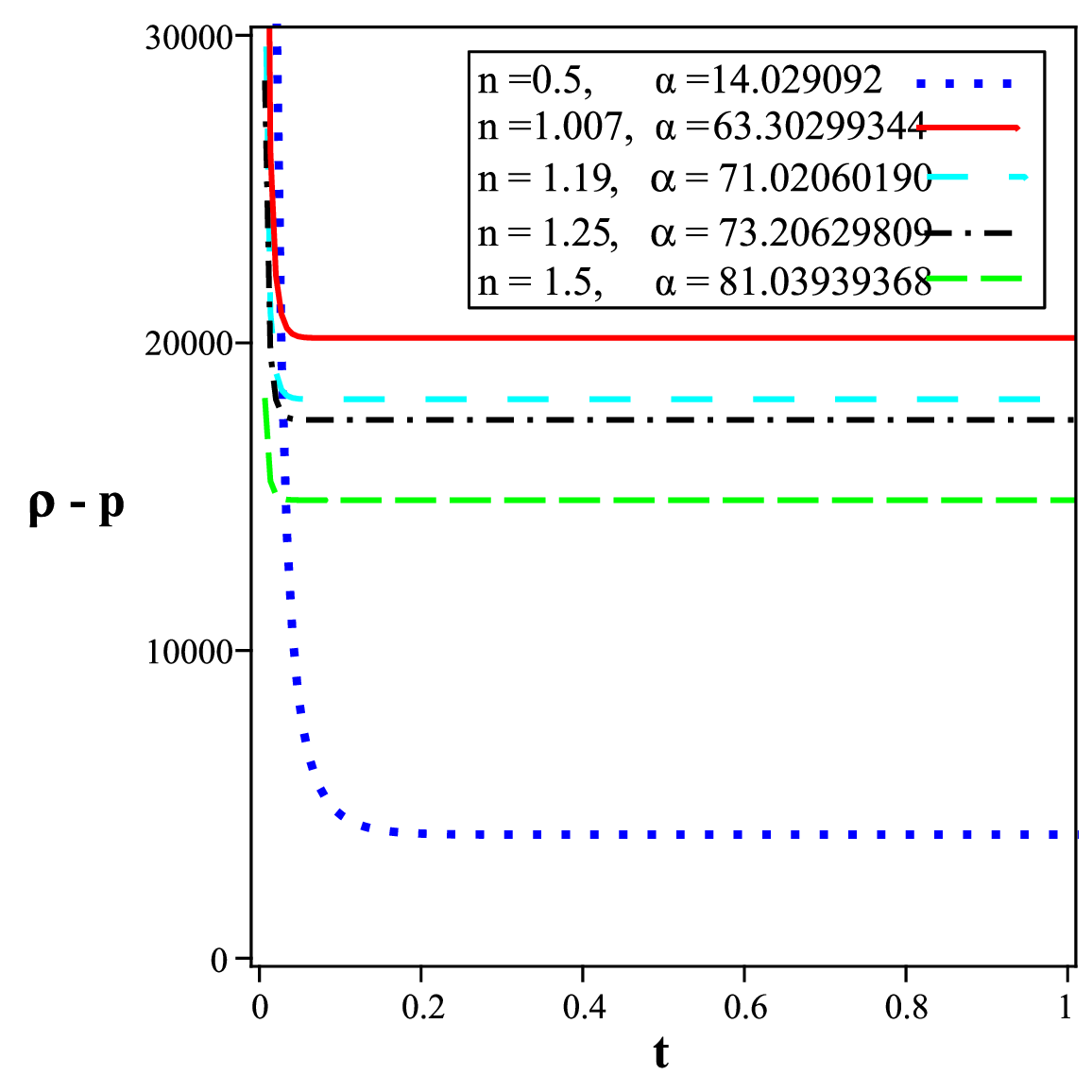}
(b)	\includegraphics[width=7.5cm,height=9cm,angle=0]{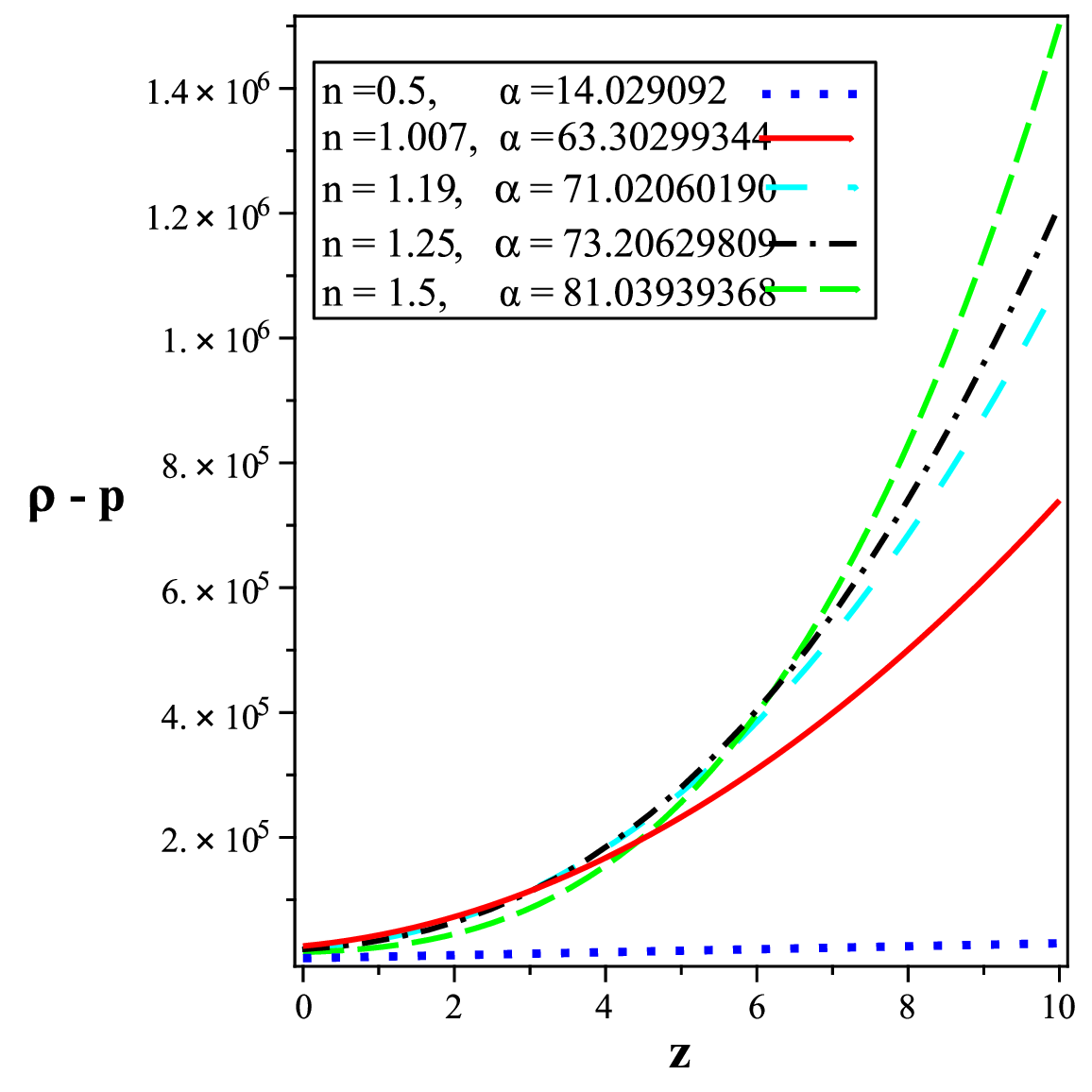}
	\caption{Plot of energy condition($\rho-p$) with cosmic time (t) in fig(a) and redshift (z) in fig(b) respectively for n= 0.5, 1.007, 1.19, 1.25, 1.5} 	
\end{figure}	
	\begin{figure}[htb]
(a)	\includegraphics[width=7.5cm,height=9cm,angle=0]{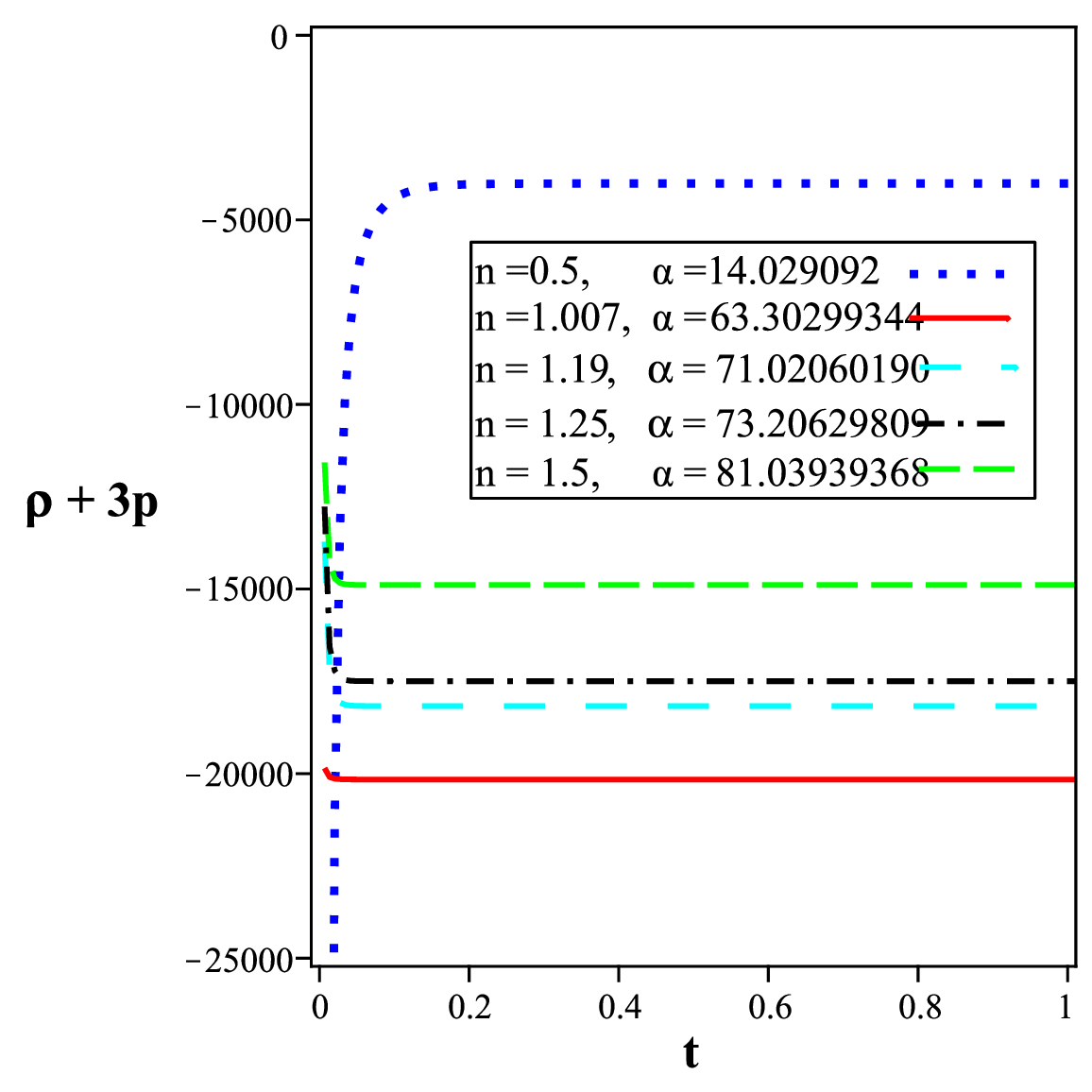}
(b)	\includegraphics[width=7.5cm,height=9cm,angle=0]{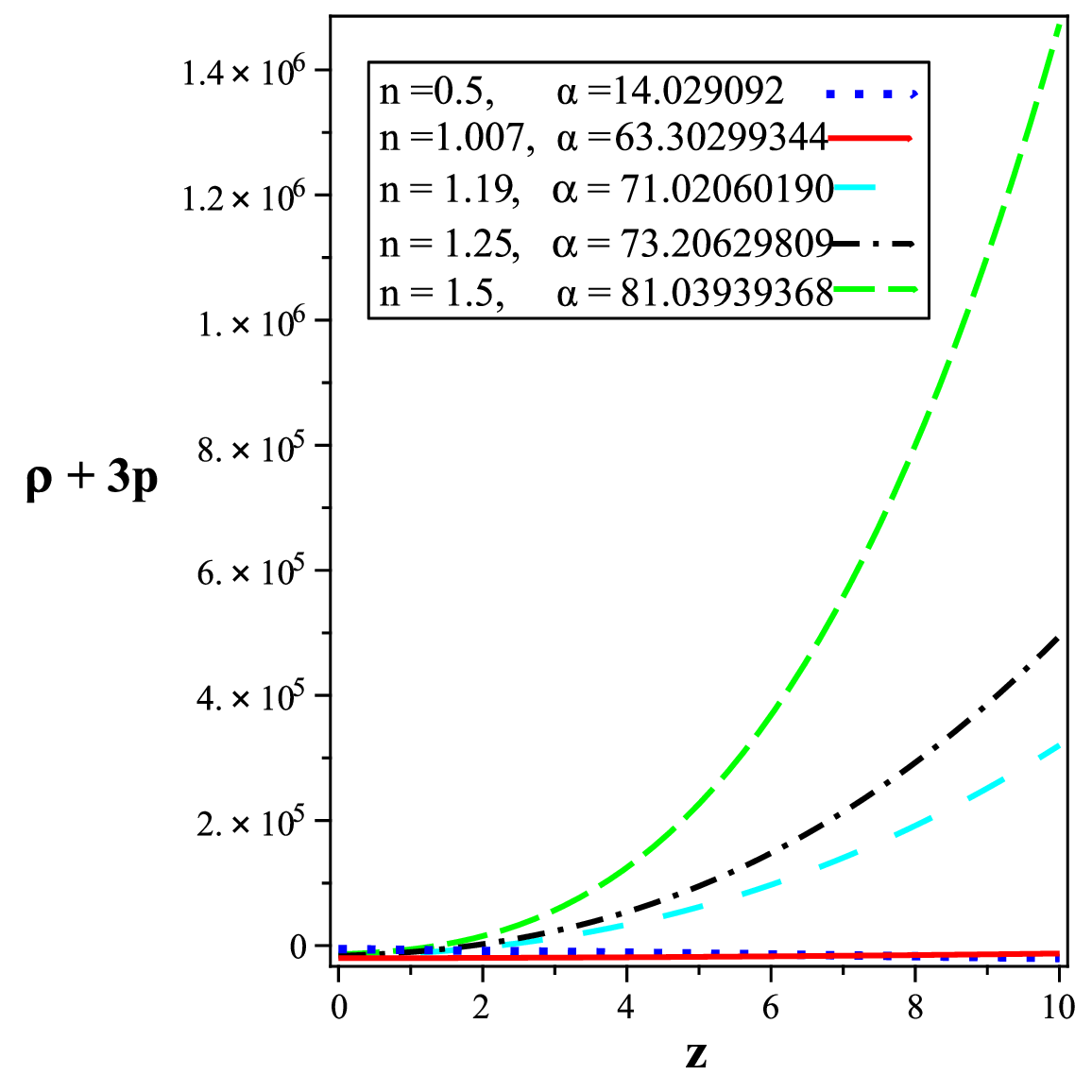}\\
	\caption{Plot of energy condition ($\rho+3p$) with cosmic time (t) in fig(a) and redshift (z) in fig(b) respectively for n= 0.5, 1.007, 1.19, 1.25, 1.5} 
\end{figure}
Figures 7(a), 8(a), and 9(a) represent the energy conditions with time (t), whereas, Figures 7(b), 8(b) , and 9(b) plots the energy conditions with redshift (z). From these figures we found that the WEC and DEC are satisfied for the derived model, and SEC is also satisfied.\\

Wald \cite{refaa54} showed that, under extreme general conditions, all Bianchi type Universes with a cosmological constant must unavoidably join a stage of exponential evolution. It has also been indicated in \cite{refaa56} that if the number of e-folds of the Universe during its exponential phase is N, then it takes a period of the order $ t\simeq e^{2N} \sqrt{\Lambda}$ for anisotropy to have an impact on the observable Universe. 

\section{Conclusion}

In this paper, we have investigated the homogeneous and anisotropic Bianchi type-V Universe filled with interacting DM and HDE. The solution of field equations has been obtained assuming a constant DP. 
We have analyzed the expansion of the spatially FLRW Universe and examined the behavior of energy density ($\rho$), pressure ($p$), the EoS parameter ($\omega$), deceleration parameter (DP), and coincidence parameter ($r$). The physical acceptability and stability of the model have been studied using the speed of sound and energy conditions.\\

We have plotted all the cosmoloical parameters for n = 0.5, 1.007, 1.19, 1.25, and 1.5. Fig. 1(a) and Fig. 1(b) show the graph of energy density versus time (t) and redshift (z) individually. The figures show that energy density is a decreasing function that converges to zero as $t \to \infty$. The plot of pressure versus cosmic time(t) and redsift (z) is presented in Fig. 2(a) and Fig. 2(b), respectively. They indicate that pressure is an increasing function. It begins with large negative value and approaches to a small negative value near zero. The EoS parameter against cosmic time (t) and redshift (z) is graphed in Figs. 3(a) and 3(b). It remains in quintessence epoch and tends to the cosmological constant ($\omega = -1$) at the future. It is fascinating to see the quintessence-like nature for the diverse estimations of n. Figures 4(a) and 4(b) presents the graph for DP versus comic time and redshift. The DP behavior has been plotted for diverse values of n. \\ 

Figures 5(a) and 5(b) show the variation of the coincidence parameter versus cosmic time and redshift at early-time. But it converges to a constant value after some limited period of time throughout the evolution. Therefore, an appropriate interaction between holographic dark energy and dark matter resembles the ratio of their densities, which is likely to acquire a fixed value under the expansion and assist in soothing the coincidence problem raised in the $\Lambda$CDM model. Figures. 6(a) and 6(b) depict the graph of the speed of sound against cosmic time and redshift. It introduces $v_{s} < 1$, throughout the evolution of the Universe. Energy conditions are plotted in Fig. 7(a) and 7(b) for $\rho+p$, for $\rho-p$ in Fig. 8(a) and 8(b), and for $\rho + 3p$ in Fig. 9(a) and 9(b). To attain the anisotropy in Bianchi type-V Universe in the background of Einstein gravity, the NEC for matter should be violated, and the EoS parameter must cross the phantom divide line ($\omega < -1$)  as shown in Fig 3(a). Strong energy condition (SEC) is negative for cosmic time, that is the case of energy violation, and the model develops in phantom phase ($\omega < -1$). According to these figures (Figs 7-9), the developed model meets the WEC and DEC.  It is also noticed that the SEC violated at the early of the cosmos, whereas it is still valid at present epoch.

\section*{Acknowledgments}
The author (A. Pradhan ) is grateful for the resources and support provided during a visit to the Inter-University Centre for Astronomy \& Astrophysics (IUCAA), Pune, India, as part of their Associateship program. The authors are grateful to the Reviewers and Editor for their insightful comments that improved the manuscript in its current form.

\section*{Conflict of Interest}
The authors declare that they have no conflicts of interest.


\end{document}